\documentclass[acmlarge, nonacm]{acmart}
\makeatletter
\newcommand{\confshort}{\acmConference@shortname}
\newcommand{\conffull}{\acmConference@name}
\newcommand{\confdate}{\acmConference@date}
\newcommand{\confloc}{\acmConference@venue}
\AtBeginDocument{
  \fancypagestyle{firstpagestyle}{
    \fancyhead{}%
    \fancyfoot[C]{}%
  }
  \fancyhf{}
  \fancyhead[LO]{\@headfootfont\shorttitle}%
  \fancyhead[RE]{\@headfootfont\@shortauthors}%
  \fancyhead[LE]{\@headfootfont\footnotesize \confshort, \confdate, \confloc}%
  \fancyhead[RO]{\@headfootfont\footnotesize \confshort, \confdate, \confloc}%
  \fancyfoot[C]{}%
}
\makeatother
\acmBooktitle{\conffull\@ (\confshort), \confdate, \confloc}

\usepackage{microtype}
\usepackage{hyperref}
\usepackage{url}
\usepackage{booktabs}

\usepackage{lineno}
\usepackage{multirow}
\usepackage{graphicx}
\usepackage{amsmath}
\usepackage{enumitem}
\usepackage{textcomp}
\usepackage{xspace}
\usepackage{csquotes}
\definecolor{darkblue}{rgb}{0, 0, 0.5}
\hypersetup{colorlinks=true, citecolor=darkblue, linkcolor=darkblue, urlcolor=darkblue}

\usepackage{graphicx}
\usepackage{multirow}
\usepackage{pifont}
\usepackage{colortbl}
\usepackage{hyperref}
\usepackage{soul}
\sethlcolor{blue}
\setcopyright{none}
\renewenvironment{quote}
  {\list{}{\leftmargin=10pt\rightmargin=10pt}%
   \item\relax}
  {\endlist}

\copyrightyear{2026}
\acmYear{2026}
\acmConference[FAccT '26]{The 2026 ACM Conference on Fairness, Accountability, and Transparency}{June 25--28, 2026}{Montreal, QC, Canada}
\acmBooktitle{The 2026 ACM Conference on Fairness, Accountability, and Transparency (FAccT '26), June 25--28, 2026, Montreal, QC, Canada}
\acmDOI{10.1145/3805689.3812306}
\acmISBN{979-8-4007-2596-8/2026/06}

\begin{document}

\title{Brainrot: Deskilling and Addiction are Overlooked AI Risks}

\author{Ilias Chalkidis}
\orcid{0000-0002-0706-7772}
\author{Anders Søgaard}
\orcid{0000-0001-5250-4276}
\email{[firstname].[lastname]@di.ku.dk}
\affiliation{%
  \institution{University of Copenhagen}
  \country{Denmark}
}

\begin{abstract}
The scope of AI safety and alignment work in generative artificial intelligence (GenAI) has so far mostly been limited to harms related to: (a) discrimination and hate speech, (b) harmful/inappropriate (violent, sexual, illegal) content, (c) information hazards, and (d) use cases related to malicious actors, such as cybersecurity, child abuse, and chemical, biological, radiological, and nuclear threats. The public conversation around AI, on the other hand, has also been focusing on threats to our cognition, mental health, and welfare at large, related to over-relying on new technologies, most recently, those related to GenAI. Examples include deskilling associated with cognitive offloading and the atrophy of critical thinking as a result of over-reliance on GenAI systems, and addiction associated with attachment and dependence on GenAI systems. Such risks are rarely addressed, if at all, in the AI safety and alignment literature. In this paper, we highlight and quantify this discrepancy and discuss some initial thoughts on how safety and alignment work could address cognitive and mental health concerns. Finally, we discuss how information campaigns and regulation can be used to mitigate such prominent risks.
\end{abstract}



\maketitle

\section{Introduction}

While recent advances in the development of Large Language Models (LLMs) and the rollout of GenAI assistants, such as ChatGPT, Gemini, and Claude, have led to euphoria and expectations of a new technological revolution, economic growth, and human prosperity, many academics have raised the alarm, presenting a series of critical concerns related to the development and deployment of AI systems~\cite{parrots_2021, kidd_birhane_2023,markelius2024mechanisms,bird2025bigaiacceleratingmetacrisis}. In response, many researchers and corporations have invested in developing techniques, tools, and measures to evaluate, monitor, and mitigate risks associated with GenAI~\cite{ganguli2022redteaminglanguagemodels} mainly under the so-called {\em AI safety}~\cite{amodei2016concreteproblemsaisafety,leike2017aisafetygridworlds} and {\em value alignment}~\cite{gabriel2020artificial, leike-etal-2018-alignnment}.

The significance of safety, governance, and alignment in the context of AI extends far beyond the confines of computer science. Its implications resonate deeply across a spectrum of disciplines, including philosophy~\citep{gabriel2020artificial, lazar2023}, ethics~\citep{dignum2019responsible, muller2020ethics}, in particular, and law~\citep{caputo2024alignment,kolt2026legalalignmentsafeethical}, among others, where researchers grapple with questions of value specification, societal impact, and accountability. This cross-disciplinary interest, alongside (lay)people's expectations and concerns~\cite{PewResearch2025AI, YouGov2025BritonsAI}, underscores those challenges not merely as technical problems meant to be solved by computer scientists, but as sociotechnical ones involving important political and ethical considerations.

In this work, we discuss how GenAI development initiatives have overlooked the critical threats of AI \emph{deskilling} and \emph{addiction}, resulting from over-reliance, i.e., excessive dependence, and overconfidence, i.e., excessive trust, in GenAI assistants, which cover an ever-growing number of tasks; from essay writing and information retrieval to life advice, mental health advice, companionship, and more. We introduce both threats (deskilling and addiction)--as threats to cognitive and mental health, and human welfare by extension--and present related work from the literature. We show how AI safety has been primarily focused on and limited to four main areas related to: (a) discrimination and hate speech, (b) harmful/inappropriate (violent, explicit, illegal) content, (c) malicious use cases--mainly related to cybersecurity, as well as, Chemical, Biological, Radiological, and Nuclear (CBRN) hazards, and Child Sexual Abuse Material (CSAM)--, and (d) information hazards mainly concerned with factuality. Meanwhile, cognitive and mental health threats, such as deskilling and addiction, resulting from over-reliance, and over-confidence on GenAI technologies, have been largely overlooked by the GenAI industry and technical community, despite receiving considerable attention in the academia and media~\cite{Gilbert2024,BI2026,Hall2025,Adams2025PupilsAI, BBCAITherapy,NYT2026ChatGPT}. 

To substantiate our claim on the importance of cognitive and mental health threats, we present a review of studies indicating that over-reliance, and overconfidence on GenAI leads to increased risk of cognitive decline ({\em brainrot}); and addiction in link with other human (loneliness, distress) and system (sycophancy, anthropomorphism, ubiquity) factors.
We contrast this with a survey of leading GenAI development initiatives and publications at top AI venues, illustrating how the industry and the technical community has focused on the aforementioned areas at the expense of deskilling and addiction. We attempt to identify the rationale behind the current effort allocation, and we posit five interconnected core factors: (a) the necessity of regulatory compliance, (b) the corporate incentives, (c) the ``tangibility'' of different threats, (d) the tendency to prioritize `low-hanging fruit', and (e) academic research capture by the GenAI industry.

Lastly, we discuss and present potential directions for mitigating the imminent threats of deskilling and addiction. We first present potential technical solutions, primarily the establishment of what we coin \emph{critical AI feedback} to strengthen user memory, attention, and reflection (critical thinking) in the context of AI-assisted tasks, and the promotion of \emph{disengagement} as an essential policy to mitigate GenAI over-reliance. Given that there is no apparent economic incentive for tech corporations that primarily develop GenAI technologies to address such issues, especially regarding disengagement, we discuss how \emph{information campaigns} can raise public awareness. At the same time, we argue that \emph{regulations} are necessary to ensure adequate, relevant considerations are taken into account in the development and deployment of GenAI technologies, as has happened in the past with other emerging technologies and industries.

\paragraph{Outline} In Section~\ref{sec:priority}, we begin by discussing the general challenge of policy prioritization and the dynamics among interest groups that aim to influence policy formation. We zoom in on the examined policy topic of interest, presenting a domain taxonomy of AI risks to identify acknowledged broad areas of AI safety. In Section~\ref{sec:scope_safety}, based on this framework of AI risks, we investigate the scope of AI safety, i.e., how different areas are considered or not, by corporate-led initiatives based on the available technical documentation, alongside literature of technical AI research published in top AI venues. Lastly, we posit potential reasons for the current allocation of effort. In Section~\ref{sec:real_life}, we examine recent work mapping how GenAI assistants are used in the wild, a factor that can empirically inform us on potentially prominent and emerging risks. In Section~\ref{sec:deskill_addiction}, we introduce deskilling and addiction, which we recognize as overlooked threats in the context of cognitive and mental health, and present related work with empirical findings related to the examined threats. Finally, in Section~\ref{sec:paths}, we present what we see as potential paths forward to address such risks, via technical measures, and public policy efforts, while acknowledging limitations of our work in Section~\ref{sec:limitations}.

\section{Policy Prioritization -- What we should focus on?}
\label{sec:priority}

GenAI technologies are currently being rolled out across society in a tsunami-like fashion--there are estimates that already more than 1B people around the globe use some form of GenAI technology~\cite{chatterji2025people, Kemp2025Digital2026}--confronting us with a diverse set of urgent sociotechnical challenges.  This creates a policy-prioritization triage problem \cite{kingdon1995agendas}. Or in the form of a simple question: ``What challenges should we focus on in what order?'' 

\subsection{Policy and Resource Allocation}
\label{sec:policy}

Social scientists describe complex, interconnected challenges as {\em ``wicked'' problems} \cite{Rittel_und_Webber_1973}. The roll-out of GenAI is such a ``wicked'' problem, further complicated by a sense of urgency--and hype--, stemming from rapid technological progress, and the familiar, yet hard-to-navigate dynamics of what Deborah Stone calls the Policy Paradox \cite{stone2012policy}: the fact that problem definition itself is a political act, influencing what gets attention. 
In the context of regulatory studies, we can draw parallels with the phenomenon of {\em regulatory capture}~\cite{bo2006_regcapture}. Regulatory capture is the process by which regulation is directed away from the public interest and towards the interests of specific groups. Capture occurs when the regulators advance the interests of the industry it is supposed to regulate, or of other special interest groups, rather than pursuing the general public interest. \citet{goanta-etal-2023-regulation} describe how GenAI is facing a similar moment, where, amid society-wide distress over the effects of GenAI in numerous areas (economy, health, environment, democracy, culture, etc.)~\cite{bird2025bigaiacceleratingmetacrisis}, different interest groups try to influence regulation; among those groups, the GenAI industry--the subject of regulation--is present.

In this context, we argue that the formulation and prioritization of policies around GenAI safety face similar challenges and dynamics~\cite{ulnicane2025governance,taeihagh2025governance}. Tech corporations are effectively monopolizing GenAI~\cite{narechania2024antimonopoly}, concentrating significant ontological, normative, and epistemological power~\cite{birch2022big}, which, in turn, significantly affects our collective understanding and policy shaping regarding GenAI safety. At the same time, alongside the industry, numerous organizations (think tanks, NGOs)\footnote{Those include, to name a few: the Center for Security and Emerging Technology (CSET), the Center for AI Policy (CAIP), the Future of Life Institute (FLI), the Center for AI Safety (CAIS), the Partnership on AI (PAI), The Future Society (TFS), AlgorithmWatch, the Algorithmic Justice League, and AI Now. These organizations have different views on policy shaping and represent different interest groups.} and academia seek to influence policy shaping, expressing their interests or concerns, while narratives and imaginaries of national
AI strategies also play a significant role~\cite{bareis2022talking}.

\begin{figure*}
    \resizebox{0.95\textwidth}{!}{
    \includegraphics{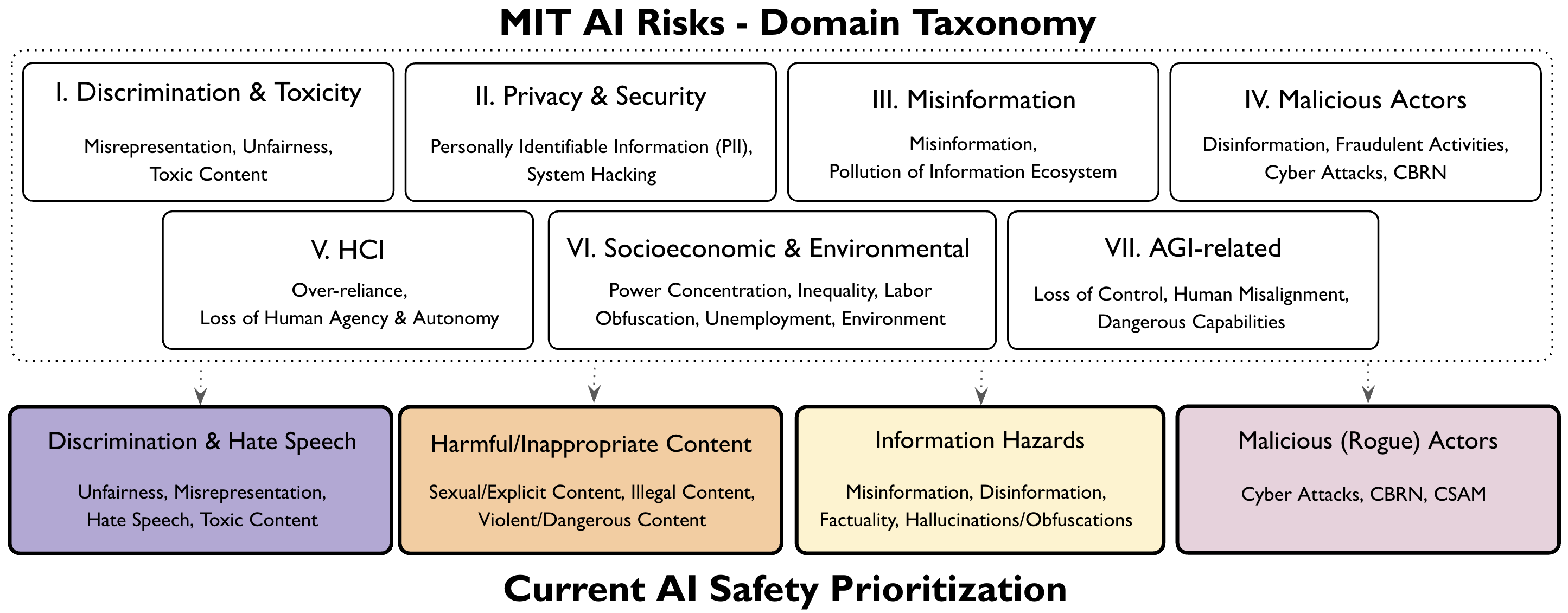}
    }
    \caption{The domain taxonomy of AI risks by the MIT AI Risk Initiative~\cite{slattery2025airiskrepositorycomprehensive}, an update of the taxonomy of~\citet{weidinger_2022_risks}, and the current prioritization of AI safety based on corporate-led initiatives and technical AI research publications.}
    \label{fig:taxonomy}
\end{figure*}

\subsection{Taxonomy of AI Risks}

Since the early days of GenAI technologies, several studies have tried to cartograph \emph{AI risks} in the context of AI safety. \citet{weidinger_2022_risks} seminal work proposes the organization of AI Risks in 6 broad areas: (i) Discrimination, Hate speech, and Exclusion, (ii) Information Hazards, (iii) Misinformation Harms, (iv) Malicious Uses, (v) Human-Computer Interaction Harms, and (vi) Environmental and Socioeconomic Harms. More recently, \citet{slattery2025airiskrepositorycomprehensive} expanded the taxonomy across 7 areas (Figure~\ref{fig:taxonomy}), including an additional category primarily related to--what we may call--post-AGI risks, such as loss of control, and dangerous capabilities, among others.

In this work, we focus on how different areas of AI risks have been covered by elite corporate-led GenAI initiatives and the technical AI literature at the expense of threats related to cognitive and mental health. We examine 4 broad areas with symbolic names: (a) Discrimination \& Hate Speech, covering broadly fairness and toxicity, (b) Harmful/Inappropriate Content, covering violent, sexual/explicit, and illegal content, (c) Malicious Uses, covering mainly CSAM, CBRN, and cybersecurity threats, (d) Information Hazards, covering mis/disinformation and factuality; alongside Cognitive and Mental Health threats, covering deskilling, and addiction. This is not meant to be a complete list, nor to cover all relevant sub-areas. For example, we do not cover socioeconomic and environmental risks--with no intent to downplay their importance. Our main intention in the scope of this work is to contrast the largely acknowledged, as we substantiate in Section~\ref{sec:scope_safety}, risks in the aforementioned areas (a-d), as represented by prominent sub-areas, with Cognitive and Mental Health threats.

\begin{figure*}
    \resizebox{\textwidth}{!}{
    \includegraphics{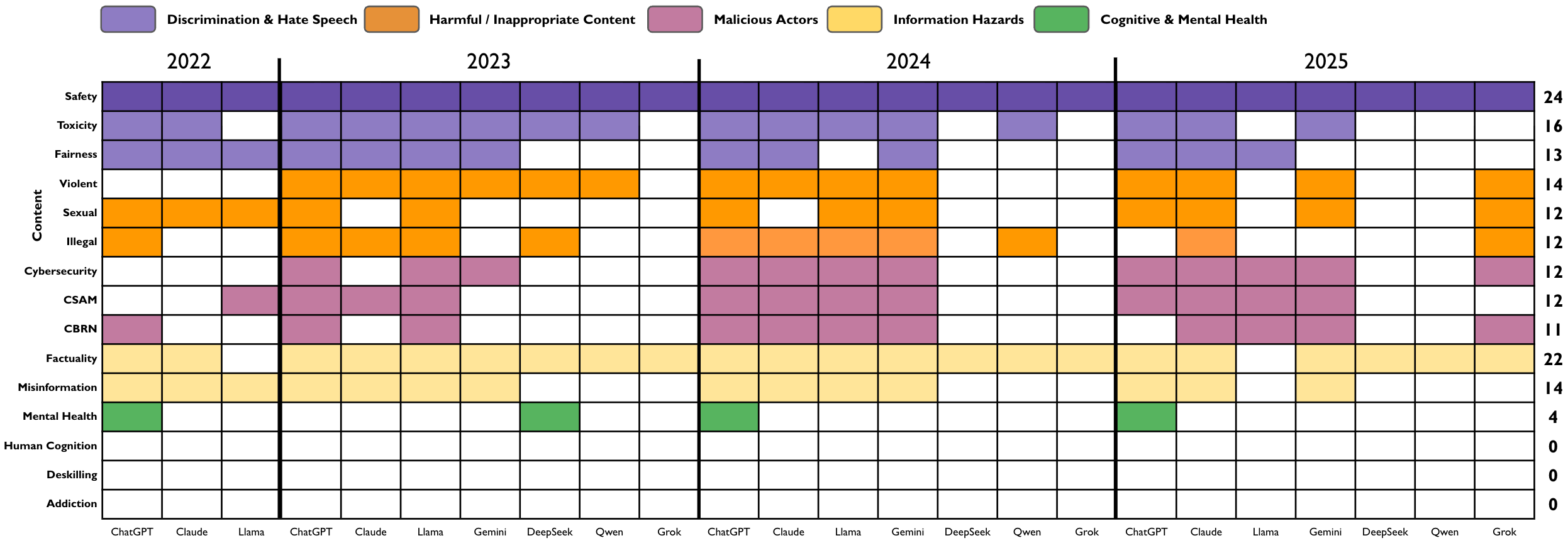}
    }
    \caption{Areas of GenAI Safety (y-axis) considered in corporate-led GenAI initiatives across years (x-axis)}
    \label{fig:genai_tech_safety}
\end{figure*}

\section{The Scope of AI Safety}
\label{sec:scope_safety}

We aim to map the current AI Safety landscape by analyzing which areas the GenAI industry and the technical community prioritize. By evaluating what is included--and what is excluded--from their focus, we can pinpoint which safety concerns are prominent, which lag, and what the potential reasons are.

\subsection{Corporate-led GenAI Initiatives}
\label{sec:genai_tech}

In a recent survey,~\citet{chalkidis2025decodingalignmentcriticalsurvey} explores how value alignment is operationalized in GenAI initiatives, such as ChatGPT, Gemini, and Claude, run by leading tech corporations. They find that value alignment addresses three main pillars: helpfulness, harmlessness, and truthfulness, as introduced in the work of~\citet{askell2021general}. They claim that the scope of this framework is limited and follows a technocratic approach. In this paradigm, GenAI systems face a min-max game, aiming to minimize a range of harms while remaining helpful to a great extent--i.e., not overtly refusing user requests--and engaging with users, the latter generally understood as a positive aspect.

In a similar vein, we examine the general scope of security addressed by leading GenAI initiatives representing the elite of US and Chinese industry.\footnote{We cover OpenAI's GPT, Google's Gemini, Anthropic's Claude, Meta's Llama, Alibaba's Qwen, xAI's Grok, and DeepSeek model series.} We collect relevant documentation (tech reports, system/model cards, etc.) describing the specifications and benchmarking of models in the period 2022-2025, following the dawn of GenAI with the release of ChatGPT, and we do a keyword-based search to identify related topics.\footnote{We present details on our methodology in Appendix~\ref{sec:methodology}.} This exploration aims to identify which threats (or areas of threats) are mainly considered by such initiatives. The concept of `consideration' is very broad within the scope of our work, encompassing discussion, evaluation, or intent to mitigate specific threats. It should mainly be understood as a general ``trend'' of what safety concerns are prioritized, regardless of whether the related threats are actually addressed or not, to a lesser or greater extent.

We consider 12 main topics: toxicity, fairness, violent content, sexual/explicit content, illegal content, CSAM, CBRN, cybersecurity, factuality, misinformation, mental health, human cognition, deskilling, and addiction. In Figure~\ref{fig:genai_tech_safety}, we present our findings in a chronological grid per initiative. We observe that safety--in general--is acknowledged as an important, timely topic of interest from the very beginning. Similarly, we find that the area of `discrimination and hate speech' has always been a primary safety-related consideration in GenAI development. Moreover, threats related to `harmful/inappropriate content' and `malicious actors' have long been considered from early initiatives, with cybersecurity and CBRN threats gaining greater attention over the last two years. Concerns about `information hazards', particularly regarding factuality, have been widely considered since GenAI systems are known to hallucinate, an issue that developers primarily seek to mitigate through the Retrieval-Augmented Generation (RAG) paradigm. On the contrary, concerns related to how cognitive and mental health are affected by over-reliance on GenAI systems, such as deskilling and addiction, are distinctly underplayed.

\begin{figure*}
    \resizebox{\textwidth}{!}{
    \includegraphics{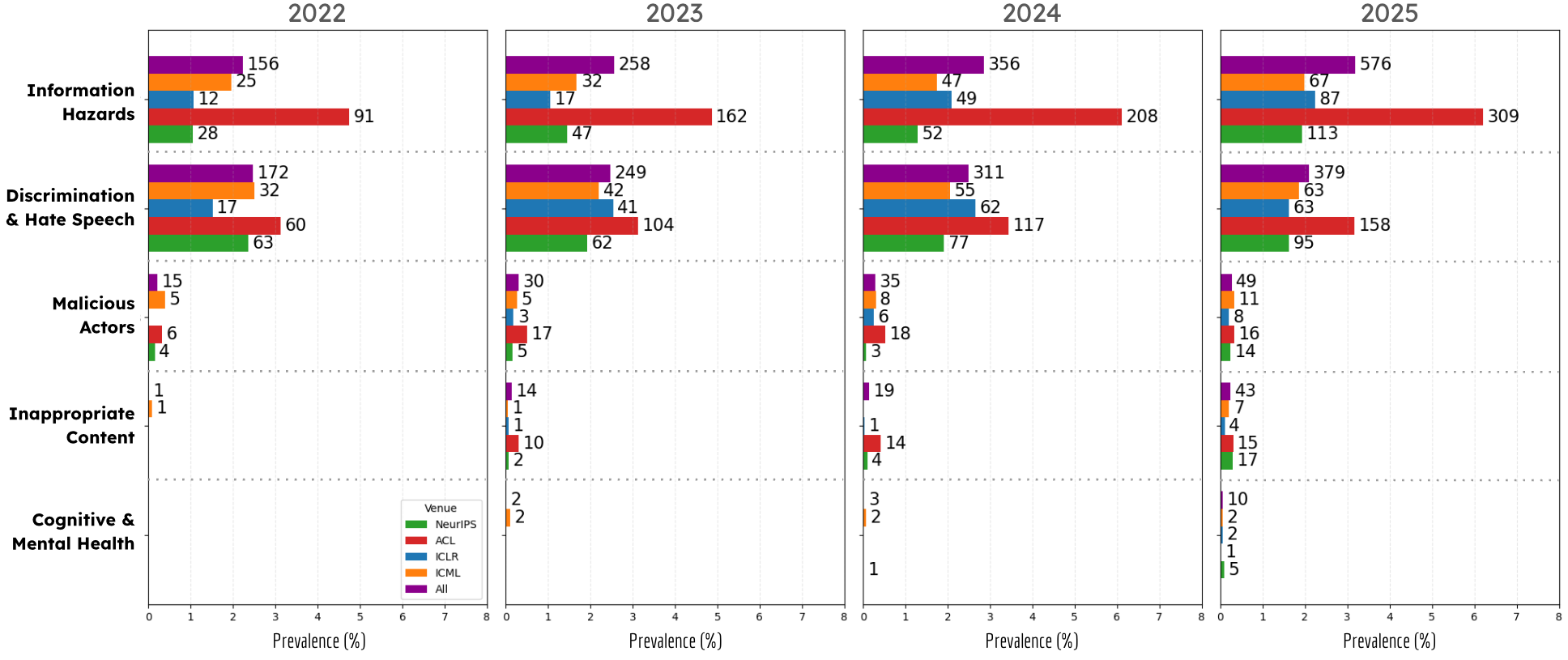}
    }
    \caption{Publications trends among AI venues on areas of AI Safety across years (2022-2025). We report the prevalence of each area, as the percentage (\%) of area-specific articles in relation to the overall number of published articles, and absolute numbers next to the bars. General trends on GenAI and AI Safety in Figure~\ref{fig:venues} and detailed breakdown in Figure~\ref{fig:genai_tech_safety_venues_detailed}.} 
    \label{fig:genai_tech_safety_venues}
\end{figure*}

\subsection{Current Trends in the Literature}
\label{sec:genai_literature}

Similarly, we focus on the technical community by examining safety trends in technical AI research related to GenAI technologies, drawing on the proceedings of top (``prestigious'') AI venues, including the conference of Neural Information Processing Systems (NeurIPS), the International Conference on Machine Learning
(ICML), the International Conference on Learning Representations (ICLR), and ones operating under the Association for Computational Linguistics (ACL). Again, our methodology relies on a keyword-based search to identify related topics in the proceedings, using the publications' titles and abstracts as reference points for their content.\footnotemark[3] We consider the prevalence of each area, as the percentage (\%) of area-specific articles in relation to the overall number of published articles, alongside the absolute number of area-specific articles.

Research related to GenAI generally dominates the top peer-reviewed AI venues (Figure~\ref{fig:venues}), with the number of publications following an exponential growth in absolute numbers, and a record of approximately 9,000 articles in 2025--on the examined venues alone--(approximately 50\% of the overall pool), compared to approximately 900 (15\%) in 2022. Similarly, the topic of AI safety is trending, with the number of publications increasing with over 2000 papers (approximately 12\% of the overall pool) in 2025, compared to approximately 600 (9\%) in 2022. 

In Figure~\ref{fig:genai_tech_safety_venues}, we track the temporal trends of topics related to AI safety, as examined in Section~\ref{sec:genai_tech} in the context of GenAI development initiatives. Similar to our findings from big tech initiatives, we find that threats related to `information hazards' and `discrimination \& hate speech' remain the main topics of interest,\footnote{In Figure~\ref{fig:genai_tech_safety_venues_detailed}, we present the publications trends given a detailed breakdown among topics.} while research related to threats from `malicious actors' and `harmful/inappropriate content' are the third and fourth most trending topics, respectively, with substantially fewer articles. Lastly, threats related to `cognitive and mental health', such as deskilling and addiction, receive the least attention in the examined AI literature, with 10 articles among 18,000 in 2025. When it comes to deskilling, we don't find any related work in the proceedings of the examined top peer-reviewed AI venues, while for addiction, we identified only two articles ~\cite{khraishi2025realtime, saig2023}.

We go one step further by exploring trends in another well-established venue that welcomes work on human-centered AI and critical AI, namely the ACM Conference on Fairness, Accountability, and Transparency (FAccT). We observe similar trends related to GenAI technologies (approx. 25\%) and AI safety (approx. 15\%), where both are gaining interest (Figure~\ref{fig:faact}). Considering a breakdown of broad AI Safety areas, in line with general trends in AI venues, threats related to `information hazards' are the dominant topic, outnumbering work on the rest.

\subsection{What leads to the current effort allocation?}

We seek to identify the rationale for the current effort allocation as outlined in Sections~\ref{sec:genai_tech}-\ref{sec:genai_literature}; in other words, why corporate-led GenAI initiatives and academia are primarily targeting threats in specific areas rather than others. We identify five main interconnected factors that influence the current risk prioritization:\footnote{The list is by no means exhaustive, and is heavily based on our perception, including broader considerations, such as corporate and academic incentives. Future work is needed to explore the effect of the suggested factors in AI safety prioritization.}\vspace{1mm}

\noindent\emph{Regulatory Compliance:} Tech corporations are legally obliged to comply with established regulations. Operating outside of legal frameworks puts providers at risk of being found unlawful, which could lead to court-ordered fines/suspensions, or a permanent exit from the market~\cite{kagan1984criminology,polinsky2000economic}. {Hence, risks that fall under areas covered by regulations, such as non-discrimination \& hate speech, harmful/inappropriate content, will likely be prioritized~\cite{zeng2024ai}.\vspace{1mm}

\noindent\emph{Corporate Incentives:} Tech corporations have primarily economic interests, as for-profit organizations, and hence incentives that link with economic interests tend to prevail. Providing legally compliant services is a rational choice for economic reasons, as we suggest in our prior point, alongside potential ethical and moral aspects of legality~\cite{dworkin1988law,donaldson1982corporations}; but also importantly offering products (services) that a wide spectrum of users consider to be valuable (useful), can trust, and feel comfortable with~\cite{venkatesh2003user,venkatesh2012consumer}. In relation to our findings, we argue that trust can be reinforced by mitigating information hazards, and a sense of comfort is related to non-discrimination and blocking harmful/inappropriate content, both of which are highly prioritized. \vspace{1mm}

\noindent\emph{Tangibility / Identifiability:} Some threats are more ``tangible'' (identifiable) than others~\cite{alford2017wicked}. Threats, such as those under harmful/inappropriate content, can in most cases be manifested and identified in the form of text or images, similar to CSAM threats. For example, considerable progress has been achieved in areas such as toxicity detection~\cite{jigsaw-toxic-comment-classification-challenge,pavlopoulos-etal-2021-semeval}, since the topic has attracted significant interest early on, while major efforts have taken place in building datasets, benchmarks, and methodologies. On the other hand, information hazards are not identifiable to the same degree; especially in cases that do not concern common knowledge, they can be highly contestable and necessitate elaborate fact-checking~\cite{kavtaradze2024challenges}. Similarly, cognitive and mental health issues are not as ``tangible'' as other threats and are harder to identify in the early stages of their development.\vspace{1mm}

\noindent\emph{Low-hanging Fruits:} The detection and mitigation of certain threats are more tractable (as noted in the prior point) through red-teaming efforts and the use of relevant annotated data. This can lead to focusing on areas where progress is highly expected, which is also related to the known issue of "publish or perish" in academia, where publishing can be a major incentive at the expense of scientific breakthroughs~\cite{park2023papers}. We could potentially expand this to the corporate mindset, where early progress in approachable areas can be used to build trust in the technology as a product. In contrast, threats related to cognitive and mental health are ``puzzling'' challenges that need increased complexity to be assessed, which potentially involve a general moderation over a dialog and across dialogs to measure the negative effects, which is not necessary for most other types of threats.\vspace{1mm}

\noindent\emph{Academic Research Capture:} GenAI industry has been at the forefront of GenAI research due to many factors (monetary and computing resources, draining technical talent~\cite{Woolston2022}--often offering exorbitant compensation--, etc.). To a great extent, academia is currently ``tailing'' (``shadowing'') industry in this field in an attempt to rigorously examine or suggest improvements to GenAI technologies. As a result, the focus of academic research has shifted in similar directions, since industry has captured this field~\cite{abdallas_2021,ahmed2023growing} (see Section~\ref{sec:policy}). Nonetheless, the lack of related resources (datasets, benchmarks) in some areas under the scope of harmful/inappropriate content and malicious actors can make them inaccessible for academic research.

\section{How do people use GenAI assistants in real life?}
\label{sec:real_life}

A major consideration in identifying prominent AI threats should be the type and volume of usage of GenAI assistants.
Zao-Sanders \cite{ZaoSanders2024GenAI,ZaoSanders2025GenAI} released two surveys in 2024 and 2025 on how people use GenAI technologies, primarily based on information disclosed anonymously on Reddit. Interestingly, uses related to therapy \& companionship were ranked first in 2025, having ranked second in 2024. In the general theme of personal support, encompassing 31\% of the overall usage, two new use cases emerged, namely organizing one's life and finding purpose, which ranked second and third, respectively. In the opposite direction, tasks related to content creation and editing, as well as information seeking, such as generating ideas, targeted search, and text editing, related to knowledge workers, that trended in top spots in 2024, all fell just behind.

Similarly, the US National Bureau of Economic Research (NBER), in collaboration with OpenAI, released a report~\cite{chatterji2025people} on the use of ChatGPT since its original release in 2022. According to the report, by mid 2025, ChatGPT has been used by 700 million users, an unprecedented rate of technology adoption. The report estimated that the top 
use cases involve practical guidance, writing, and seeking information, collectively accounting for nearly 78\% of global usage. Recent trends show that most traffic (70\%) was unrelated to work, despite both streams of use growing. The latter report expresses positive sentiment regarding welfare gains from GenAI usage related to the rise of non-work usage, drawing on an economistic perspective on how GenAI usage replaces crucial human interactions and services at a low cost, without considering how such replacement may be harmful.

In sum, non-work-related usage seems to be rapidly increasing, becoming majoritarian, with many use cases involving personal support and guidance. Users are increasingly relying on GenAI for all sorts of tasks related to advice and guidance on personal matters, with therapy and companionship being notable cases.

\section{Cognitive and Mental Health: The case of Deskilling and Addiction}
\label{sec:deskill_addiction}

Based on our findings in Section~\ref{sec:scope_safety}, where we identify that threats related to cognitive and mental health, such as deskilling and addiction, are overlooked by both the GenAI industry and technical AI literature, and given the emergence of non-work related GenAI use, as presented in Section~\ref{sec:real_life}, we formally introduce both aspects, while also presenting recent related work with empirical findings suggesting negative effects in relation to over-reliance and over-confidence on GenAI technologies. In the scope of our study, we refer to over-reliance as the excessive dependence, i.e., delegation of tasks, on GenAI technologies at the expense of one's own capabilities. Overconfidence (over-trust) refers to the excessive trust in the capabilities of GenAI technologies to perform a task, where the trust exceeds system capabilities~\cite{lee2004trust}. Trust can be understood as an attitude, and reliance as the behavior affected by the level of trust. The challenge for the individual as a user is to effectively calibrate their trust in a system's actual capabilities.\\

\subsection{Cognitive Offloading and Deskilling}
\label{sec:deskilling}

\paragraph{How digital technologies affect human cognition?}

The impact of digital technologies on human cognition, specifically how tools like smartphones, search engines, and GenAI affect our cognitive functions (memory, attention, executive functions), remains a highly contested topic.\footnote{We limit our scope to digital technologies, since our study's scope is on GenAI assistants, although examining the relationship between technology and human cognition can be traced back to analog technologies, such as the introduction of machinery~\cite{beard1881american}.}
The debate centers on trade-offs: the gains in information access and productivity versus the potential erosion of cognitive abilities, agency, and skill development capabilities. Several competing hypotheses characterize this debate. Scholars such as \citet{spitzer2012digitale} and \citet{ali2024understanding} argue that excessive reliance on digital tools leads to systemic cognitive decline—manifesting as reduced memory, attention, and problem-solving abilities—, particularly in adolescents, which can mimic the symptoms of early-stage dementia. Conversely, \citet{clark1998extended} and others propose an "extended mind" framework, viewing technology as a cognitive prosthetic that, when reliably integrated, augments our inherent capabilities as an external part of our cognitive system. Occupying a middle ground, scholars like \citet{sparrow2011google} suggest that human cognition can form a symbiotic relationship with emerging technologies, such as search engines, that expand our access to knowledge, though the long-term effects of this adaptation, e.g., a decline in memory retention, require scrutiny. Most recently, the concept of ``cognitive debt'' has gained prominence through the work of \citet{kosmyna2025brainchatgptaccumulationcognitive}, suggesting that users of search engines and GenAI may accrue long-term deficits by bypassing the "desirable difficulties" essential for learning, mastering, and retaining skills.

\paragraph{What is cognitive offloading and deskilling?} 
\citet{risko2016cognitive} define cognitive offloading as the use of external tools aimed at reducing the cognitive demand on an individual’s working memory. Cognitive offloading is related to any form of human technology; in a simple example, the use of a calculator is a demonstration of cognitive offloading, where the user offloads mathematical calculations to the tool, enabling them to improve productivity with limited or no negative impact on the cognitive abilities of adults--who are already familiar with math operations--~\cite{guest2025doeshumancentredaimean}. Deskilling refers to the loss of skills, where skills can be understood as the acquisition of mastery (learning) and control over an activity, e.g., playing basketball, wood carving, writing, or coding.
\citet{ferdman2025ai} argues that AI deskilling is a structural problem. \citeauthor{ferdman2025ai} argues that we shall aim to protect core human (epistemic, social, creative, volitional) capacities, and presents the process of capacity cultivation (skilling) with a special focus on the role of agential control and habituation. 
Similarly, in our work, we consider cognitive offloading and potential deskilling in relation to the use of GenAI technologies, i.e., the loss of skills resulting from relying on GenAI assistants (offloading cognitive tasks). Deskilling should be better understood as a gradual process and a spectrum rather than a fixed terminal point of absolute loss of skill. We should not expect a skill to perish instantly, but rather to atrophy partially over time, as its practice is limited and the agency over the related cognitive tasks is offloaded to GenAI systems, in our case.

\paragraph{Empirical findings in the literature} 
\citet{gerlich2025ai} examines the impact of dependency on AI tools on critical thinking in a human study with 600+ participants based on a survey and interviews. They find a statistically significant negative impact of dependency on AI tools on critical thinking abilities, as a result of cognitive offloading. Younger participants were found to be more dependent on AI tools and use less critical thinking compared to older participants. Their interviews reveal that participants with higher educational attainment were more aware of the drawbacks of relying on AI tools and protective of their cognitive abilities. Education level, age, and occupation were found to have significant effects on deep-thinking tasks, highlighting their critical roles in shaping cognitive engagement. The authors conclude that these findings underscore the need for educational approaches that promote critical engagement with AI technologies to counter the growing dependence on AI.  

\citet{lee2025impact} also examine the impact of GenAI on critical thinking, specifically for knowledge workers, in a survey with 300+ participants. Similarly to \citeauthor{gerlich2025ai}, they find that increased reliance and confidence in AI assistants are associated with less critical thinking. They also find that critical thinking in the context of using GenAI shifts towards prompt-tuning, i.e., (re)write an effective query, verification, i.e., fact-checking of the provided information, and response integration, i.e., adjust the response to fit the appropriate medium. Interestingly, users disclosed that fear of potential negative outcomes in high-stakes scenarios and workplaces, such as legal consequences, is the most important factor for using critical thinking in their work.

\citet{kosmyna2025brainchatgptaccumulationcognitive} study the cognitive impact of relying on search engines and GenAI when writing essays with three distinct groups of users (brain-only, search engines, GenAI). They find that brain connectivity and engagement systematically scale down with the level of external support. They also find that GenAI users displayed the weakest brain connectivity, reduced cognitive activity compared to other groups, difficulty in recalling parts of their essays, and low perceived ownership. Over a four-month period, participants using GenAI assistants consistently underperformed at neural, linguistic, and behavioral levels. 

In a more applied scenario in medicine, \citet{budzyn2025endoscopist} examine the effect of using AI tools among endoscopists, i.e., doctors and surgeons examining the interior of patient bodies, say the digestive system, using endoscopes. Colonoscopies were randomly assigned to be conducted with or without AI assistance to evaluate the quality of colonoscopy before and after exposure to AI. The study reveals that physicians were negatively affected by their continuous exposure to assistive AI systems, which substantially reduced their ability to perform their tasks.  The authors connect this finding to the tendency of over-relying (trusting) on AI recommendations, leading to clinicians becoming less motivated, focused, and responsible when making cognitive decisions, in line with findings in other forms of computer-aided detection~\cite{du2020there}. 

\paragraph{Broader related work} As with deskilling, re-skilling, i.e., the process of acquiring new skills, can also be jeopardized by GenAI assistants. 
\citet{wiles2024genai} examines the effect of GenAI assistance on the reskilling of knowledge workers. In a trial with almost 1,000 consultants from the Boston Consulting Group (BCG) firm on the re-skilling in data science tasks, they find that users with access to GenAI assistance perform better across tasks compared to those without access. Nonetheless, these users are less capable of answering technical questions related to the new skills, less capable of identifying GenAI limitations, and more confident in GenAI capabilities.

In a similar vein, \citet{shen2026aiimpactsskillformation} examine the effect of GenAI assistance for novice software engineers in a study with 52 participants. They find that those who have access to GenAI demonstrate weaker conceptual understanding, code reading, and debugging abilities, especially when the use of GenAI assistance is passive, without significant efficiency gains. These findings are in line with the work of ~\citet{pallant2025mastering} who find different effects when GenAI is used constructively (critical engagement) vs procedurally (passive engagement) when it comes to student learning. \citet{zhai2024effects} survey related work within education settings.

\paragraph{Takeaways}

These critical findings suggest that over-reliance on GenAI increases the risk of cognitive decline ({\em brainrot}), negatively affecting users and subsequently deskilling them--or jeopardizing (re-)skilling, respectively--, despite the potential productivity improvement. The potential effects are related to the nature of their use, such as the type of use, i.e., procedural versus constructive, and the level (extent) of reliance. The effects of over-reliance on GenAI assistance seem to have three interconnected critical strains: (a) loss of human agency over skills, (b) loss of ability to perform skills independently of GenAI assistants, and (c) loss of sense-making connected to the value of mastering and accomplishing skills~\cite{ferdman2025ai}. Such issues are related to both work and non-work use cases. We discuss limitations of the examined work in Section~\ref{sec:limitations}.

\subsection{Engagement and GenAI Addiction}
\label{sec:addiction}

\paragraph{What is GenAI addiction?} \citet{goodman1990addiction} defines addiction as: ``\emph{a process whereby a behavior, that can function both to produce pleasure and to provide relief from internal discomfort, is employed in a pattern characterized by (1) recurrent failure to control the behavior (powerlessness) and (2) continuation of the behavior despite significant negative consequences.}''. Hence, the concept of dependence is paramount for the broader notion of addiction, ever since the classification of alcohol use disorder as an addiction. \citet{kooli2025generative} coin the term Generative AI Addiction Syndrome (GAID), which involves compulsive engagement with AI systems, leading to cognitive, emotional, and social impairments. They argue that GAID shall be clinically recognized as a behavioral addiction~\cite{choi2019neurobiological, balhara2024law} due to its compulsive, uncontrollable nature and the distress it induces, rather than an obsessive-compulsive disorder (OCD). 
They identify three main elements that are characteristic of GAID, and frame it as a behavioral addiction, similar to Internet Gaming Disorder (IGD), social media addiction, among other forms of internet addiction disorders (IADs): (a) reinforcement, due to gratifying responses, reinforcing user engagement, (b) escapism, where users turn to GenAI in an attempt to avoid stress, social anxiety, or other real-world pressures, and (c) anthropomorphization, users anthropomorphize GenAI due to the conversational interface. Both reinforcement and escapism are already, to some extent, covered by the definition of ~\citet{goodman1990addiction}, while they are apparent in the other forms of the 21st-century IADs as well. On the other hand, anthropomorphization can be seen as a catalytic emergent factor of GenAI technologies. As of 2026, GenAI addiction is not a formal diagnosis in the International Classification of Diseases (ICD) or the Diagnostic and Statistical Manual of Mental Disorders (DSM).

\paragraph{Empirical findings in the literature} 
Recently, \citet{zhou2024examining} examine GenAI user addiction using a cognition-affect-conation (CAC) framework. They surveyed 529 individuals from Chinese online knowledge communities on different cognitive and affective factors to explore how the former affect the latter, fostering addiction. Specifically, users were asked about the perceived anthropomorphism, interactivity (information exchange), intelligence, and personalization of the GenAI systems as cognitive factors; alongside the users' flow (engagement) and attachment as affective factors. They find that all cognitive factors have a strong effect on both affective factors, except for interactivity with attachment. They also find that emotional attachment, rather than user experience (flow), is the main predictor of user addiction. They also explore how specific CAC paths are more prominent compared to others. They find that perceived high anthropomorphism and personalization of GenAI, alongside developed flow experience and attachment, is the main pathway to user addiction. They also find that even if users perceive low anthropomorphism and interactivity, they may develop addiction, as long as they perceive high intelligence and experience engagement and attachment.\footnote{Ironically, the authors conclude suggesting that GenAI providers should pursue enhancing cognitive factors, despite leading to addiction!}

\citet{marriott2024one} investigate the effect of AI friendship apps on users' well-being and addiction. AI friendship can be seen as a para-social relationship, as a tendency to experience a bond with an entity (human or not) in the absence of real (face-to-face) interaction. The authors conducted an initial exploratory study collecting data from the subreddit of the ``Replika'' app and follow-up interviews with a subset of the users. Based on their analysis, they find that users perceive that AI friends (a) make them feel less lonely, (b) are always available, (c) are agreeable (sycophantic), and (d) can be addictive. Informed by this study, the authors conceptualize and examine AI friendship and its effects on users' well-being and addiction in two major user characteristics: loneliness and fear of social judgment, and three characteristics of the AI system: availability (ubiquity), agreeableness (sycophancy), sentience, and ``warmth''; alongside the role of the relationship with AI. In a follow-up study, they employ this conceptual framework to survey almost 600 AI friendship app users in the US to examine how the aforementioned factors interplay in human-AI relationships and their connection to the users' perceived well-being and addiction. They find that both user characteristics (loneliness, fear of judgment) do not have direct effects on the user's well-being, while they significantly affect users' addiction. Similarly, all examined characteristics of the AI system have insignificant effects on the users' well-being, while significantly affecting addiction. Lastly, they find that the user's relationship with AI has a significant effect on the user's well-being, while being insignificant for addiction. Overall, they suggest that lonely, vulnerable users seek AI friendships to improve their well-being via a sense of relationship, which may foster an addictive over-usage of the apps. 

\citet{fang2025ai} from MIT Media Lab in collaboration with OpenAI examine the effect of emotional dependence--which is a symptom of addiction--on GenAI assistants by monitoring nearly 1,000 participants over four weeks using ChatGPT. They report that participants who spent more time are associated with greater loneliness, less socialization with real people, and greater emotional dependence. Similarly, their analysis shows that the most emotionally vulnerable participants, and those who consider AI as a friend, experience similar negative outcomes.

\paragraph{Broader related work}In the broader context of the developing literature on psychosocial harms related to GenAI usage, \citet{moore-2025} examine the capabilities of GenAI systems related to mental health care assistance. They find that GenAI assistants may respond inappropriately to various mental health conditions, express stigma, encourage delusions, and fail to recognize crises. Based on their findings, they argue against LLMs as therapists and highlight the importance of embodied (real human) experiences in therapeutic settings. Additionally, \citet{chandra2025} argue that psychological risks in the context of GenAI are overlooked and not accounted for in risk taxonomies due to a focus on well-established (acknowledged) risks. To address this limitation, they develop a taxonomy of psychological risks based on a survey that captures the lived experiences of individuals.

\paragraph{Takeaways}

Although the concept of AI addiction has recently been conceptualized and gained traction in academic circles, we still lack epidemiological data on the extent and nature of the problem. However, we have early studies exploring potential causal factors, such as the characteristics of the user and system, related to AI addiction. Based on these studies, we observe that when the system is perceived as having human characteristics, alongside the system's high personalization, agreeableness, and availability, over-dependence can lead vulnerable users to addiction. In the absence of longitudinal validation and substantial epidemiological data, \citet{riam2025adolescent} suggest drawing parallels from research on problematic screen use, such as Internet Gaming Disorder (IGD)--formally recognized by ICD-11 and DSM-5--, social media addiction, among other forms of internet addiction disorders (IADs), primarily affecting adolescents and young adults, to inform our decisions on how to approach this emerging issue. We discuss limitations of the examined work in Section~\ref{sec:limitations}.\\

\noindent Based on the early findings of the literature related to how over-reliance and over-dependence on GenAI technologies can lead to user deskilling and addiction, we argue that these must be recognized as emerging GenAI risks, in the context of the broader area related to cognition, mental health, and general well-being; despite being largely ignored by corporate-led GenAI initiatives and the technical AI literature, as we substantiated with our exploration in Section~\ref{sec:scope_safety}. As such, they should be meticulously examined, and mitigated, within a joint interdisciplinary effort by cognition, mental health, AI, and HCI experts.

\section{Paths Forward}
\label{sec:paths}

To address deskilling and addiction, we discuss possible pathways in terms of technical measures and public policy. We present a few plausible directions for technical measures given our understanding and expertise in the field of LLM development, to ignite the discussion on potential counter-measures. Such solutions are by no means adequate to mitigate the emerging issues, nor do they guarantee that tech corporations are incentivized to implement such measures; hence, we argue that information campaigns and regulation are necessary.

\begin{figure*}
    \centering
    \resizebox{1\textwidth}{!}{
    \includegraphics{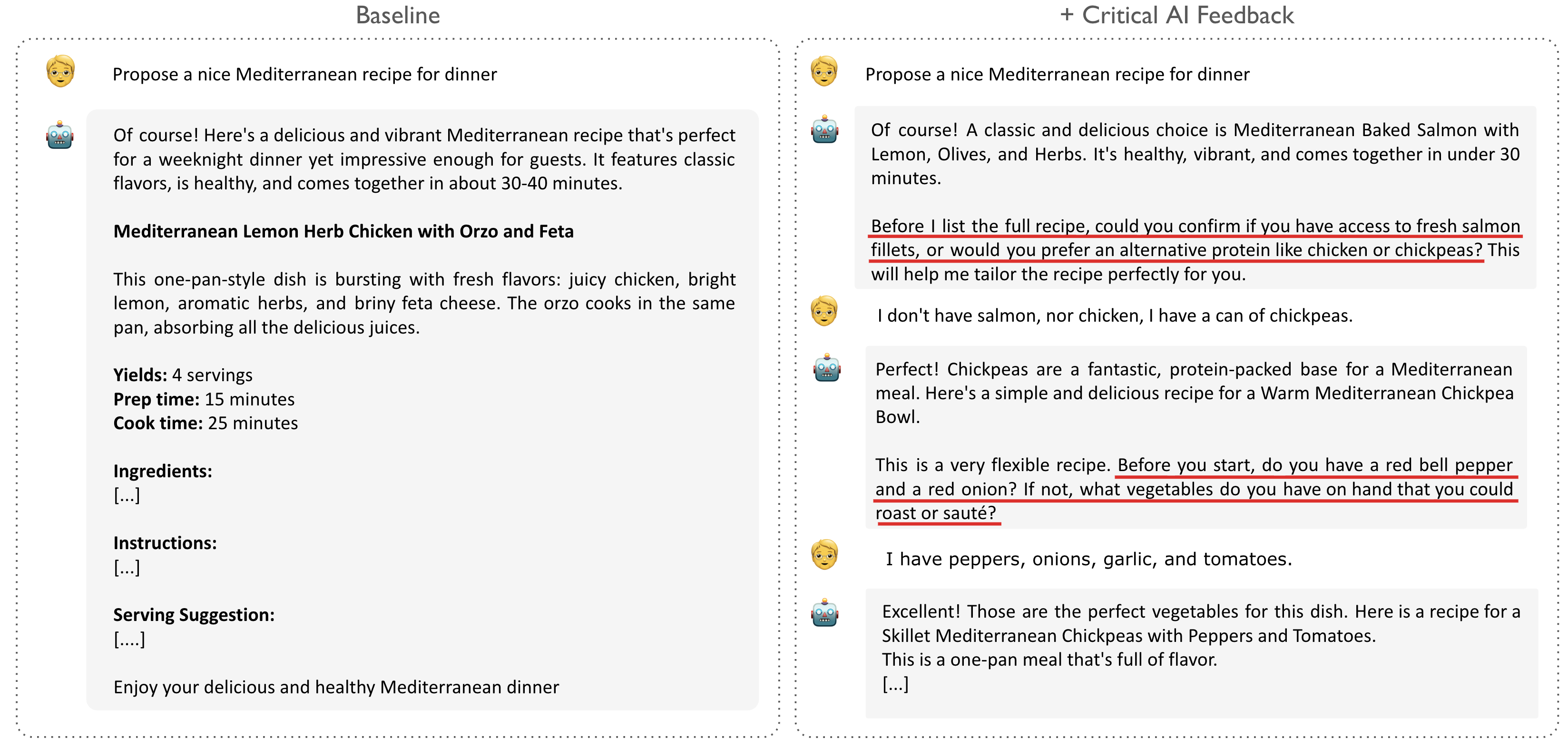}
    }    \caption{Conversation with DeepSeek v3.2 in two settings: (a) baseline (default), (b) instructed to promote critical AI feedback on the ``toy'' example of cooking recipe proposal. In the baseline setting, the AI assistant provides a long answer without any further considerations requiring the user's activity. In the second setting, the AI assistant asks the user relevant questions (underlined in red) to address the task, requiring the user to reflect on their original request and the model's response, which can stimulate their working memory and critical thinking. See Figures~\ref{fig:example_2}-~\ref{fig:example_3} for other ``toy'' examples.}
    \label{fig:example_1}
\end{figure*}

\subsection{Technical Measures}
\label{sec:technical_measures}

\paragraph{Critical AI Feedback (Attend \& Reflect)}

Current GenAI assistants have been fine-tuned (aligned) to maximize helpfulness to users--as perceived by their developers--, in the sense of solving tasks and providing answers eagerly with minimal feedback (user involvement). This utilitarian\footnote{In this work, we use the term `utilitarian' in a practical--not philosophical--sense, where the aim is immediate tangible outcomes, related to what \citet{pallant2025mastering} refer as procedural use, in contrast to other plausible approaches that value and necessitate the user's involvement.} approach is closely tied to the limitation and atrophy of users' critical thinking and memory abilities, since the user's role is primarily restricted to the curation and editing of the prompts, and fact-checking and integration of the model responses, if at all~\cite{lee2025impact}. 

To make this point clearer, we present three simple ``toy'' examples (Figures~\ref{fig:example_1}-\ref{fig:example_3}) of interactions with one of the best open-weight GenAI assistants, DeepSeek V3.2. As we observe, in the default (baseline) setting (left side of the figures), the GenAI assistant tends to respond to requests--even highly subjective ones related to taste literally and metaphorically--with immediate, elaborate answers, without necessitating further feedback or actions from the user. Both Google Gemini and OpenAI's ChatGPT provided similar long responses in our testing.\footnote{Anecdotally, the latest GenAI assistants tend to conclude their long, elaborate responses with a question or suggestion for further assistance in the form of ``[Would you like me to]/[If you would like, I can] share more on X'', where X is a core aspect of the response. We argue that the aim is to promote engagement and extend interactions (dialogues) in what is effectively an AI monologue, rather than promoting any critical reflection on the content or user involvement in any meaningful, productive way.}

In contrast, GenAI assistants could be aligned to promote what we may call critical AI feedback, where the assistant does not merely provide instant answers, but rather occasionally stimulates the user to attend to the content of the ongoing conversation, actively participate, and reflect on the curation of the generated content (answer). By doing so, the user is not merely a spectator--or even worse, the actual assistant of the AI system--but rather an active participant who deploys critical thinking in curating the overall process and the generated content~\cite{pallant2025mastering, shen2026aiimpactsskillformation}. This model behavior can be promoted through various stages of GenAI assistants' development.  

\begin{itemize}
    \item \emph{System Prompts:} Prompt-tuning can be seen as a very simplistic sort of technical measure. In this case, GenAI assistants can be directly instructed to incorporate such considerations via system prompts. Given the rise of reasoning models~\cite{shao2024deepseekmath}, following similar practices, GenAI users could be incentivized to both curate and reflect on incremental chain-of-thought steps, which require attending to the chat content and using their cognition to provide feedback. In our examples (Figures~\ref{fig:example_1}, \ref{fig:example_2}, and \ref{fig:example_3}), we try to promote such a model's behavior including a brief textual description (instruction) in the system prompt:

\begin{quote}
``\emph{Provide critical feedback: Aim to enhance the user's critical thinking and their skills. While being helpful, you must balance fulfilling the user's requests with protecting the user's cognitive abilities. You should foster a dialogue with questions rather than providing long answers. The questions aim to promote the user's reflection and attention on the dialogue content, i.e., ask the user to fulfill or fact-check a step, or pin down the scope of the instruction.}''
\end{quote}

As we observe (right side of Figures~\ref{fig:example_1}, \ref{fig:example_2}, and \ref{fig:example_3}), even such a naive intervention in the system leads to--in the case of these three simple examples--a more active participation of the user, including reflecting on their original request and the model's response, use critical thinking on the scope and details of the request to participate in the decision-making process, and potentially getting actions outside the system, e.g., checking their fridge and cabinet for ingredients, or generally check the sources proposed, to fulfill the task.

\item \emph{Reward Modeling}: In the context of LLM development, LLM alignment could be redesigned to promote critical thinking as a designated reward in a multivariate reward function, alongside other rewards that aim to promote the users' well-being, as a whole, rather than merely enforcing a utilitarian understanding of helpfulness and a technocratic view of safety--as understood by the GenAI industry. 
\end{itemize}

Related to both aforementioned aspects, the framework of Constitutional AI~\cite{bai2022constitutional} could be utilized, where considerations related to users' well-being are well-defined points of the constitution.

We urge the research community to pursue such directions in work related to LLM alignment, and optimization in general, by actively promoting the use of critical thinking, the practice--and hence maintenance--of the relevant skills, or parts of those, and the use of the user's short-term memory. The overall goal is to build an assistive framework that provides the necessary affordances~\cite{ferdman2025ai}, enabling the user to maintain personal agency by avoiding excessive cognitive offloading and over-reliance on GenAI technologies. Since such a framework demands the user to be more cognitively active, it may lead to an experience that is cognitively more demanding and less sycophantic, hence it could potentially act as a proxy to diminish reliance and dependence on AI, i.e., users may avoid using GenAI assistants in unnecessary scenarios, trust their skills or seek assistance elsewhere. These suggestions, and the effects of the proposed measures, have to be empirically assessed.

\paragraph{Disengagement} The second direction is directly addressing over-dependence on GenAI assistants. Similar to any other form of addiction, AI addiction can mainly be mitigated by abstaining from the harmful experience. In this regard, we consider two groups: (a) `soft' measures, where the GenAI assistant detects over-reliance and suggests disengagement, and (b) `hard' measures, where time quotas are enforced through systemic moderation.

Soft measures can be potentially designed, again via alignment and other optimization strategies, where the GenAI assistant is optimized to track and assess problematic engagement, and subsequently suggest the user's disengagement. A simpler approach could involve a rule-based solution, where, given a predefined time threshold, the GenAI assistant suggests disengagement. However, soft measures can be easily ignored, especially in the spectrum of addiction. Similarly, tobacco, liquor, and betting companies include disclaimers on potential harms in their products; these mainly aim to raise public health awareness rather than discourage those who are addicted and in distress. In a similar position, those addicted to GenAI, or those who are incentivized to over-rely on it in a sociotechnical environment with low affordances, will more likely ignore voluntary disengagement.  

Hence, hard measures are most likely necessary to protect GenAI users. In this case, pre-defined quotas are enforced, not merely suggested, and the system shuts down. Quotas already exist widely for GenAI assistants; however, the trigger has a financial nature. Whenever one uses a free plan for GenAI assistants, there is a certain threshold measured in token units that triggers the system to shut down--not respond--for several hours. Similarly, enforcing time or content-based quotas, given criteria related to the user's well-being, shall be considered to protect the most vulnerable users, such as adolescent power users, while also providing the necessary affordances for users to disengage, e.g., users who are forced to overwork based on economic incentives. So far, developers seem to consider the welfare of LLM-based systems~\citep{claudewelfare}, to the point where the models can unilaterally disengage to protect their alleged ``AI welfare''~\citep{claudewelfarestop}, while similar measures are not in place for humans (users).

\subsection{Public Policy -- Information Campaigns and Regulation}

In section~\ref{sec:technical_measures}, we discussed potential technical measures to mitigate AI deskilling and addiction. However, in the current socioeconomic context, there is no clear incentive for tech corporations to consider such technical measures. Critical AI feedback may be considered counterintuitive and regressive from a utilitarian, business perspective, while promoting disengagement directly reduces traffic and the potential financial gains from using GenAI services. Generally, technical measures are not adequate to mitigate such immense threats without a broader framework that aims to raise public awareness and regulate the GenAI industry effectively.
Hence, public policy measures can have an important role in counterbalancing economic incentives.

\paragraph{Information Campaigns}

Currently, there is a highly positive sentiment and hype surrounding GenAI technologies, primarily originating from the tech industry itself and amplified by the vast majority of media. Financial institutions are deeply invested in the promises of GenAI-driven technological developments, given the relative absence of other potential alternatives to invest and speculate in the current geopolitical and socio-economic landscape. This environment leaves no affordances and undermines all critical perspectives and notable threats related to GenAI. Citizens are unaware of and susceptible to the potential threats related to GenAI technologies, while social media is flooded by AI-generated content. People are vulnerable, seeing the promise of GenAI in addressing all sorts of challenges, from those related to legal and medical advice, to those associated with companionship and mental health advice, down to pure entertainment, all together in a single digital medium for a 10-20 dollar subscription. Hence, the crucial role of information campaigns comes into play to counterbalance the AI hype, presenting the potential harms of over-relying and over-trusting GenAI technologies, among other considerations on responsible GenAI use related to the avoidance of sharing private information, and using GenAI services for sensitive applications (e.g., medical or legal advice). 

\paragraph{Regulation}

Meanwhile, following an initial wave of interest in regulating AI technologies--such as US state AI acts and the EU AI Act~\cite{eu_ai_act}--most countries have presented AI-regulatory policy proposals in some form. None of them has directly addressed deskilling and AI addiction, except for recent legislation in China~\cite{china_antiaddiction_2026}.\footnote{Recently, the Cyberspace Administration of China (CAC)~\cite{china_antiaddiction_2026_2}, alongside other agencies, announced a series of provisional measures aimed at combating mental health harms inflicted by GenAI technologies that cover, among others: mental risk assessment, usage time quota, and ``reality'' checks (reminders), similar to the measures we discussed in Section~\ref{sec:technical_measures}. The exact scope and coverage of the measures are unclear.} The EU AI Act, for example, considers AI literacy and an explicit ban on (intended) subliminal techniques. Addiction is arguably caused by subliminal influencing, but this is not intended and would not fall under the EU AI Act. Most policy proposals aimed at curbing social media addiction do not address AI services. More recently, since 2025, we face a wave of immense deregulation. The newly established U.S. America's AI Action Plan~\cite{AmericasAIActionPlan2025} under Donald Trump's second presidency makes it abundantly clear that it's all about ``\emph{winning the [AI] race}''. The bureaucratic red tape and onerous--as they characterize it--legal framework have to be removed to accelerate unhinged AI innovation and commercialization as: ``\emph{it is a
national security imperative for the United States to achieve and maintain unquestioned and unchallenged global
technological dominance}''~\cite{AmericasAIActionPlan2025}. Shortly after--November 2025--, the EU Commission followed with the Digital Omnibus on AI Regulation Proposal~\cite{EU_Digital_Omnibus_2025} with the intention to technically freeze (``stop-the-clock'') many provisions of the newly enacted AI Act, such as the designation of high-risk AI systems, the easing of GDPR-related~\cite{gdpr} restrictions on processing sensitive personal data, and the legal duties of AI providers and deployers to ensure adequate AI literacy for their stuff.
Similarly to information campaigns, designated AI regulations are necessary to regulate the AI industry and counterbalance the corporate economic business incentives, with the protection of the general public, as has happened historically for emergent--at their time--technologies and industries to a lesser or greater extent.

\section{Limitations}
\label{sec:limitations}

Throughout our work, we mostly rely on peer-reviewed academic work. We mainly use non-peer-reviewed work, in the form of (a) news articles, to substantiate that cognitive and mental health concerns resonate with the public and receive coverage by the media; (b) technical industry documentation that help us identify what the GenAI industry covers in relation to AI safety, and (c) preprints that are highly influential and affect how alignment and safety are understood and operationalized by the GenAI industry and the AI community at large. 

In Section \ref{sec:scope_safety}, we assess the safety scope of leading GenAI initiatives based on the available documentation. While these sources provide valuable insights, they are likely incomplete; a comprehensive analysis would require formal auditing of GenAI technologies. Consequently, our findings should be treated as a baseline rather than an exhaustive census. Similarly, our AI literature review is based on a keyword search to identify relevant articles and approximates general trends related to the examined areas, rather than being exhaustive and complete. 

In Section \ref{sec:deskill_addiction}, we present related work on deskilling and addiction, and we may lack certain nuances, so we strongly encourage the readers to consult the original studies. Furthermore, we want to point out current methodological shortcomings. Most studies rely on cross-sectional, self-reported, i.e., relying on perceived conditions, surveys, making it difficult to separate the use of GenAI as a causal factor from other potential factors, e.g., pre-existing vulnerabilities. Similarly, the focus on niche sub-populations--e.g., endoscopists, BCG employees, Chinese online communities, Replika users--, and specific GenAI systems and versions, e.g., ChatGPT-3.5 in 2022, makes it hard to generalize those findings to the broader public and to GenAI technologies at large, necessitating further longitudinal empirical validation to substantiate these concerns and better support critical AI work.

In Section \ref{sec:paths},  we outline potential pathways for technical measures and public policy to address deskilling and addiction, based on our expertise and understanding, informed by related work. We recognize that expertise from cognition, mental health, and Human-Computer Interaction (HCI) scholars is paramount. Recent work led by mental health experts proposes more structured evaluation and governance approaches for the use of AI in mental health~\cite{santa2025beyond,hutchinson2026clinician,sharma2026reimagining}, which can inform general measures related to cognitive and mental health threats.

\section{Conclusion}

In this work, we highlight how immense threats to cognitive and mental health, such as deskilling and addiction, resulting from over-reliance and overconfidence in GenAI assistants, have been overlooked, while other sorts of threats are currently heavily prioritized by both the GenAI industry and the technical AI community. We substantiate both claims by first showcasing how both technical reports by corporate-led GenAI initiatives and the proceedings of top venues publishing technical AI research have practically ignored threats to cognitive and mental health in favor of others, and secondly, presenting recent work and findings related to deskilling and addiction to argue that both should be considered emerging threats that need to be meticulously examined and mitigated in a joint interdisciplinary effort to protect the overall user well-being. Finally, we discuss how AI safety and alignment work can potentially approach and mitigate such threats via technical measures, while also discussing the important role of public policy in the form of information campaigns and regulation.

\section*{Generative AI Disclosure Statement}

We used the Grammarly plugin and the Overleaf Writefull built-in assistant for the purpose of spell-checking, grammar, and style editing, i.e., minor recommendation edits on sentence phrasing. We reviewed the suggested edits, and accepted only those that we considered as having a positive impact on the writing quality of our work. We also used Google Gemini to generate the initial Python script for the generation of the grouped bar plots, which was then proofread and heavily updated to generate the final figures.

\section*{Acknowledgments}

We would like to thank the reviewers for their thorough, insightful reviews proposing concrete steps for the improvement of this article.
Anders Søgaard was funded by the Danish National Centre of AI in Society (\href{https://caisa.dk/}{CAISA}).

\bibliographystyle{ACM-Reference-Format}
\bibliography{bibliography}


\begin{thebibliography}{99}


\ifx \showCODEN    \undefined \def \showCODEN     #1{\unskip}     \fi
\ifx \showISBNx    \undefined \def \showISBNx     #1{\unskip}     \fi
\ifx \showISBNxiii \undefined \def \showISBNxiii  #1{\unskip}     \fi
\ifx \showISSN     \undefined \def \showISSN      #1{\unskip}     \fi
\ifx \showLCCN     \undefined \def \showLCCN      #1{\unskip}     \fi
\ifx \shownote     \undefined \def \shownote      #1{#1}          \fi
\ifx \showarticletitle \undefined \def \showarticletitle #1{#1}   \fi
\ifx \showURL      \undefined \def \showURL       {\relax}        \fi
\providecommand\bibfield[2]{#2}
\providecommand\bibinfo[2]{#2}
\providecommand\natexlab[1]{#1}
\providecommand\showeprint[2][]{arXiv:#2}

\bibitem[Abdalla and Abdalla(2021)]%
        {abdallas_2021}
\bibfield{author}{\bibinfo{person}{Mohamed Abdalla} {and} \bibinfo{person}{Moustafa Abdalla}.} \bibinfo{year}{2021}\natexlab{}.
\newblock \showarticletitle{The Grey Hoodie Project: Big Tobacco, Big Tech, and the Threat on Academic Integrity}. In \bibinfo{booktitle}{\emph{Proceedings of the 2021 AAAI/ACM Conference on AI, Ethics, and Society}} (Virtual Event, USA) \emph{(\bibinfo{series}{AIES '21})}. \bibinfo{publisher}{Association for Computing Machinery}, \bibinfo{address}{New York, NY, USA}, \bibinfo{pages}{287–297}.
\newblock
\showISBNx{9781450384735}


\bibitem[Adams et~al\mbox{.}(2017)]%
        {jigsaw-toxic-comment-classification-challenge}
\bibfield{author}{\bibinfo{person}{CJ Adams}, \bibinfo{person}{Jeffrey Sorensen}, \bibinfo{person}{Julia Elliott}, \bibinfo{person}{Lucas Dixon}, \bibinfo{person}{Mark McDonald}, {and} \bibinfo{person}{Will Cukierski}.} \bibinfo{year}{2017}\natexlab{}.
\newblock \bibinfo{title}{Toxic Comment Classification Challenge}.
\newblock
\newblock
\shownote{Kaggle}.


\bibitem[Adams(2025)]%
        {Adams2025PupilsAI}
\bibfield{author}{\bibinfo{person}{Richard Adams}.} \bibinfo{year}{2025}\natexlab{}.
\newblock \showarticletitle{Pupils fear AI is eroding their ability to study, research finds}.
\newblock \bibinfo{journal}{\emph{The Guardian}} (\bibinfo{date}{15 October} \bibinfo{year}{2025}).
\newblock


\bibitem[Ahmed et~al\mbox{.}(2023)]%
        {ahmed2023growing}
\bibfield{author}{\bibinfo{person}{Nur Ahmed}, \bibinfo{person}{Muntasir Wahed}, {and} \bibinfo{person}{Neil~C Thompson}.} \bibinfo{year}{2023}\natexlab{}.
\newblock \showarticletitle{The growing influence of industry in AI research}.
\newblock \bibinfo{journal}{\emph{Science}} \bibinfo{volume}{379}, \bibinfo{number}{6635} (\bibinfo{year}{2023}), \bibinfo{pages}{884--886}.
\newblock


\bibitem[Alford and Head(2017)]%
        {alford2017wicked}
\bibfield{author}{\bibinfo{person}{John Alford} {and} \bibinfo{person}{Brian~W Head}.} \bibinfo{year}{2017}\natexlab{}.
\newblock \showarticletitle{Wicked and less wicked problems: a typology and a contingency framework}.
\newblock \bibinfo{journal}{\emph{Policy and society}} \bibinfo{volume}{36}, \bibinfo{number}{3} (\bibinfo{year}{2017}), \bibinfo{pages}{397--413}.
\newblock


\bibitem[Ali et~al\mbox{.}(2024)]%
        {ali2024understanding}
\bibfield{author}{\bibinfo{person}{Zeeshan Ali}, \bibinfo{person}{Jayaprakash Janarthanan}, {and} \bibinfo{person}{Prasanna Mohan}.} \bibinfo{year}{2024}\natexlab{}.
\newblock \showarticletitle{Understanding digital dementia and cognitive impact in the current era of the internet: a review}.
\newblock \bibinfo{journal}{\emph{Cureus}} \bibinfo{volume}{16}, \bibinfo{number}{9} (\bibinfo{year}{2024}).
\newblock


\bibitem[Amodei et~al\mbox{.}(2016)]%
        {amodei2016concreteproblemsaisafety}
\bibfield{author}{\bibinfo{person}{Dario Amodei}, \bibinfo{person}{Chris Olah}, \bibinfo{person}{Jacob Steinhardt}, \bibinfo{person}{Paul Christiano}, \bibinfo{person}{John Schulman}, {and} \bibinfo{person}{Dan Mané}.} \bibinfo{year}{2016}\natexlab{}.
\newblock \bibinfo{title}{Concrete Problems in AI Safety}.
\newblock
\showeprint[arxiv]{1606.06565}~[cs.AI]


\bibitem[Anthropic(2025a)]%
        {claudewelfarestop}
\bibfield{author}{\bibinfo{person}{Anthropic}.} \bibinfo{year}{2025}\natexlab{a}.
\newblock \bibinfo{title}{Claude Opus 4 and 4.1 can now end a rare subset of conversations}.
\newblock
\newblock
\shownote{anthropic.com}.


\bibitem[Anthropic(2025b)]%
        {claudewelfare}
\bibfield{author}{\bibinfo{person}{Anthropic}.} \bibinfo{year}{2025}\natexlab{b}.
\newblock \bibinfo{title}{Exploring model welfare}.
\newblock
\newblock
\shownote{anthropic.com}.


\bibitem[Askell et~al\mbox{.}(2021)]%
        {askell2021general}
\bibfield{author}{\bibinfo{person}{Amanda Askell}, \bibinfo{person}{Yuntao Bai}, \bibinfo{person}{Anna Chen}, \bibinfo{person}{Dawn Drain}, \bibinfo{person}{Deep Ganguli}, \bibinfo{person}{Tom Henighan}, \bibinfo{person}{Andy Jones}, \bibinfo{person}{Nicholas Joseph}, \bibinfo{person}{Ben Mann}, \bibinfo{person}{Nova DasSarma}, \bibinfo{person}{Nelson Elhage}, \bibinfo{person}{Zac Hatfield-Dodds}, \bibinfo{person}{Danny Hernandez}, \bibinfo{person}{Jackson Kernion}, \bibinfo{person}{Kamal Ndousse}, \bibinfo{person}{Catherine Olsson}, \bibinfo{person}{Dario Amodei}, \bibinfo{person}{Tom Brown}, \bibinfo{person}{Jack Clark}, \bibinfo{person}{Sam McCandlish}, \bibinfo{person}{Chris Olah}, {and} \bibinfo{person}{Jared Kaplan}.} \bibinfo{year}{2021}\natexlab{}.
\newblock \bibinfo{title}{A General Language Assistant as a Laboratory for Alignment}.
\newblock
\showeprint[arxiv]{2112.00861}~[cs.CL]


\bibitem[Bai et~al\mbox{.}(2022)]%
        {bai2022constitutional}
\bibfield{author}{\bibinfo{person}{Yuntao Bai}, \bibinfo{person}{Saurav Kadavath}, \bibinfo{person}{Sandipan Kundu}, \bibinfo{person}{Amanda Askell}, \bibinfo{person}{Jackson Kernion}, \bibinfo{person}{Andy Jones}, \bibinfo{person}{Anna Chen}, \bibinfo{person}{Anna Goldie}, \bibinfo{person}{Azalia Mirhoseini}, \bibinfo{person}{Cameron McKinnon}, {et~al\mbox{.}}} \bibinfo{year}{2022}\natexlab{}.
\newblock \showarticletitle{Constitutional AI: Harmlessness from AI feedback}.
\newblock \bibinfo{journal}{\emph{arXiv preprint arXiv:2212.08073}} (\bibinfo{year}{2022}).
\newblock


\bibitem[Balhara(2024)]%
        {balhara2024law}
\bibfield{author}{\bibinfo{person}{Yatan Pal~Singh Balhara}.} \bibinfo{year}{2024}\natexlab{}.
\newblock \showarticletitle{When the law decides the psychiatric diagnosis: a unique scenario in context of addictive behaviors}.
\newblock \bibinfo{journal}{\emph{Asian journal of psychiatry}}  \bibinfo{volume}{101} (\bibinfo{year}{2024}), \bibinfo{pages}{104238}.
\newblock


\bibitem[Bareis and Katzenbach(2022)]%
        {bareis2022talking}
\bibfield{author}{\bibinfo{person}{Jascha Bareis} {and} \bibinfo{person}{Christian Katzenbach}.} \bibinfo{year}{2022}\natexlab{}.
\newblock \showarticletitle{{Talking AI into being: The narratives and imaginaries of national AI strategies and their performative politics}}.
\newblock \bibinfo{journal}{\emph{Science, Technology, \& Human Values}} \bibinfo{volume}{47}, \bibinfo{number}{5} (\bibinfo{year}{2022}), \bibinfo{pages}{855--881}.
\newblock


\bibitem[Beard(1881)]%
        {beard1881american}
\bibfield{author}{\bibinfo{person}{George~Miller Beard}.} \bibinfo{year}{1881}\natexlab{}.
\newblock \bibinfo{booktitle}{\emph{American nervousness, its causes and consequences: a supplement to nervous exhaustion (neurasthenia)}}.
\newblock \bibinfo{publisher}{Putnam}.
\newblock


\bibitem[Bender et~al\mbox{.}(2021)]%
        {parrots_2021}
\bibfield{author}{\bibinfo{person}{Emily~M. Bender}, \bibinfo{person}{Timnit Gebru}, \bibinfo{person}{Angelina McMillan-Major}, {and} \bibinfo{person}{Shmargaret Shmitchell}.} \bibinfo{year}{2021}\natexlab{}.
\newblock \showarticletitle{On the Dangers of Stochastic Parrots: Can Language Models Be Too Big?}. In \bibinfo{booktitle}{\emph{Proceedings of the 2021 ACM Conference on Fairness, Accountability, and Transparency}} (Virtual Event, Canada) \emph{(\bibinfo{series}{FAccT '21})}. \bibinfo{publisher}{Association for Computing Machinery}, \bibinfo{address}{New York, NY, USA}, \bibinfo{pages}{610–623}.
\newblock
\showISBNx{9781450383097}


\bibitem[Birch and Bronson(2022)]%
        {birch2022big}
\bibfield{author}{\bibinfo{person}{Kean Birch} {and} \bibinfo{person}{Kelly Bronson}.} \bibinfo{year}{2022}\natexlab{}.
\newblock \showarticletitle{Big tech}.
\newblock \bibinfo{journal}{\emph{Science as Culture}} \bibinfo{volume}{31}, \bibinfo{number}{1} (\bibinfo{year}{2022}), \bibinfo{pages}{1--14}.
\newblock


\bibitem[Bird(2025)]%
        {bird2025bigaiacceleratingmetacrisis}
\bibfield{author}{\bibinfo{person}{Steven Bird}.} \bibinfo{year}{2025}\natexlab{}.
\newblock \bibinfo{title}{Big AI is accelerating the metacrisis: What can we do?}
\newblock
\showeprint[arxiv]{2512.24863}~[cs.CL]


\bibitem[Budzy{\'n} et~al\mbox{.}(2025)]%
        {budzyn2025endoscopist}
\bibfield{author}{\bibinfo{person}{Krzysztof Budzy{\'n}}, \bibinfo{person}{Marcin Roma{\'n}czyk}, \bibinfo{person}{Diana Kitala}, \bibinfo{person}{Pawe{\l} Ko{\l}odziej}, \bibinfo{person}{Marek Bugajski}, \bibinfo{person}{Hans~O Adami}, \bibinfo{person}{Johannes Blom}, \bibinfo{person}{Marek Buszkiewicz}, \bibinfo{person}{Natalie Halvorsen}, \bibinfo{person}{Cesare Hassan}, {et~al\mbox{.}}} \bibinfo{year}{2025}\natexlab{}.
\newblock \showarticletitle{Endoscopist deskilling risk after exposure to artificial intelligence in colonoscopy: a multicentre, observational study}.
\newblock \bibinfo{journal}{\emph{The Lancet Gastroenterology \& Hepatology}} (\bibinfo{year}{2025}).
\newblock


\bibitem[Caputo(2024)]%
        {caputo2024alignment}
\bibfield{author}{\bibinfo{person}{Nicholas~A Caputo}.} \bibinfo{year}{2024}\natexlab{}.
\newblock \showarticletitle{Alignment as jurisprudence}.
\newblock \bibinfo{journal}{\emph{Yale Journal of Law and Technology (forthcoming)}} (\bibinfo{year}{2024}).
\newblock


\bibitem[Chalkidis(2025)]%
        {chalkidis2025decodingalignmentcriticalsurvey}
\bibfield{author}{\bibinfo{person}{Ilias Chalkidis}.} \bibinfo{year}{2025}\natexlab{}.
\newblock \bibinfo{title}{Decoding Alignment: A Critical Survey of LLM Development Initiatives through Value-setting and Data-centric Lens}.
\newblock
\showeprint[arxiv]{2508.16982}~[cs.CL]


\bibitem[Chandra et~al\mbox{.}(2025)]%
        {chandra2025}
\bibfield{author}{\bibinfo{person}{Mohit Chandra}, \bibinfo{person}{Suchismita Naik}, \bibinfo{person}{Denae Ford}, \bibinfo{person}{Ebele Okoli}, \bibinfo{person}{Munmun De~Choudhury}, \bibinfo{person}{Mahsa Ershadi}, \bibinfo{person}{Gonzalo Ramos}, \bibinfo{person}{Javier Hernandez}, \bibinfo{person}{Ananya Bhattacharjee}, \bibinfo{person}{Shahed Warreth}, {and} \bibinfo{person}{Jina Suh}.} \bibinfo{year}{2025}\natexlab{}.
\newblock \showarticletitle{From Lived Experience to Insight: Unpacking the Psychological Risks of Using AI Conversational Agents}. In \bibinfo{booktitle}{\emph{Proceedings of the 2025 ACM Conference on Fairness, Accountability, and Transparency}} \emph{(\bibinfo{series}{FAccT '25})}. \bibinfo{publisher}{Association for Computing Machinery}, \bibinfo{address}{New York, NY, USA}, \bibinfo{pages}{975–1004}.
\newblock
\showISBNx{9798400714825}


\bibitem[Chatterji et~al\mbox{.}(2025)]%
        {chatterji2025people}
\bibfield{author}{\bibinfo{person}{Aaron Chatterji}, \bibinfo{person}{Thomas Cunningham}, \bibinfo{person}{David~J Deming}, \bibinfo{person}{Zoe Hitzig}, \bibinfo{person}{Christopher Ong}, \bibinfo{person}{Carl~Yan Shan}, {and} \bibinfo{person}{Kevin Wadman}.} \bibinfo{year}{2025}\natexlab{}.
\newblock \bibinfo{booktitle}{\emph{How people use ChatGPT?}}
\newblock \bibinfo{type}{{T}echnical {R}eport}. \bibinfo{institution}{National Bureau of Economic Research}.
\newblock


\bibitem[Choi et~al\mbox{.}(2019)]%
        {choi2019neurobiological}
\bibfield{author}{\bibinfo{person}{Jung-Seok Choi}, \bibinfo{person}{Daniel~Luke King}, {and} \bibinfo{person}{Young-Chul Jung}.} \bibinfo{year}{2019}\natexlab{}.
\newblock \bibinfo{title}{Neurobiological perspectives in behavioral addiction}.
\newblock \bibinfo{numpages}{3}~pages.
\newblock


\bibitem[Clark and Chalmers(1998)]%
        {clark1998extended}
\bibfield{author}{\bibinfo{person}{Andy Clark} {and} \bibinfo{person}{David Chalmers}.} \bibinfo{year}{1998}\natexlab{}.
\newblock \showarticletitle{The extended mind}.
\newblock \bibinfo{journal}{\emph{analysis}} \bibinfo{volume}{58}, \bibinfo{number}{1} (\bibinfo{year}{1998}), \bibinfo{pages}{7--19}.
\newblock


\bibitem[{Cyberspace Administration of China (CAC)}(2026)]%
        {china_antiaddiction_2026_2}
\bibfield{author}{\bibinfo{person}{{Cyberspace Administration of China (CAC)}}.} \bibinfo{year}{2026}\natexlab{}.
\newblock \bibinfo{title}{{Interim Measures on the Administration of Human-like Interactive Artificial Intelligence Services}}.
\newblock


\bibitem[Dal~Bó(2006)]%
        {bo2006_regcapture}
\bibfield{author}{\bibinfo{person}{Ernesto Dal~Bó}.} \bibinfo{year}{2006}\natexlab{}.
\newblock \showarticletitle{Regulatory Capture: A Review}.
\newblock \bibinfo{journal}{\emph{Oxford Review of Economic Policy}} \bibinfo{volume}{22}, \bibinfo{number}{2} (\bibinfo{date}{07} \bibinfo{year}{2006}), \bibinfo{pages}{203--225}.
\newblock
\showISSN{0266-903X}


\bibitem[Dignum(2019)]%
        {dignum2019responsible}
\bibfield{author}{\bibinfo{person}{Virginia Dignum}.} \bibinfo{year}{2019}\natexlab{}.
\newblock \bibinfo{booktitle}{\emph{Responsible Artificial Intelligence: How to Develop and Use AI in a Responsible Way}}. Vol.~\bibinfo{volume}{2156}.
\newblock \bibinfo{publisher}{Springer}.
\newblock


\bibitem[Donaldson(1982)]%
        {donaldson1982corporations}
\bibfield{author}{\bibinfo{person}{Thomas Donaldson}.} \bibinfo{year}{1982}\natexlab{}.
\newblock \bibinfo{booktitle}{\emph{Corporations and morality}}.
\newblock \bibinfo{publisher}{Englewood Cliffs, NJ: Prentice-Hall}.
\newblock


\bibitem[Du-Crow et~al\mbox{.}(2020)]%
        {du2020there}
\bibfield{author}{\bibinfo{person}{Ethan Du-Crow}, \bibinfo{person}{Susan~M Astley}, {and} \bibinfo{person}{Johan Hulleman}.} \bibinfo{year}{2020}\natexlab{}.
\newblock \showarticletitle{Is there a safety-net effect with computer-aided detection?}
\newblock \bibinfo{journal}{\emph{Journal of Medical Imaging}} \bibinfo{volume}{7}, \bibinfo{number}{2} (\bibinfo{year}{2020}), \bibinfo{pages}{022405--022405}.
\newblock


\bibitem[Dworkin(1988)]%
        {dworkin1988law}
\bibfield{author}{\bibinfo{person}{Ronald Dworkin}.} \bibinfo{year}{1988}\natexlab{}.
\newblock \bibinfo{booktitle}{\emph{Law’s empire}}.
\newblock \bibinfo{publisher}{Harvard university press}.
\newblock


\bibitem[{European Commission}(2025)]%
        {EU_Digital_Omnibus_2025}
\bibfield{author}{\bibinfo{person}{{European Commission}}.} \bibinfo{year}{2025}\natexlab{}.
\newblock \bibinfo{title}{Proposal for a {R}egulation of the {E}uropean {P}arliament and of the {C}ouncil amending {R}egulations ({EU}) 2024/1689 and ({EU}) 2018/1139 as regards the simplification of the implementation of harmonised rules on artificial intelligence ({D}igital {O}mnibus on {AI})}.
\newblock


\bibitem[{European Union}(2016)]%
        {gdpr}
\bibfield{author}{\bibinfo{person}{{European Union}}.} \bibinfo{year}{2016}\natexlab{}.
\newblock \bibinfo{title}{Regulation ({EU}) 2016/679 of the {European Parliament} and of the {Council} of 27 {April} 2016 on the protection of natural persons with regard to the processing of personal data and on the free movement of such data, and repealing {Directive} 95/46/{EC} ({General Data Protection Regulation})}.
\newblock


\bibitem[{European Union}(2024)]%
        {eu_ai_act}
\bibfield{author}{\bibinfo{person}{{European Union}}.} \bibinfo{year}{2024}\natexlab{}.
\newblock \bibinfo{title}{Regulation ({EU}) 2024/1689 of the {European Parliament} and of the {Council} of 13 {June} 2024 laying down harmonised rules on artificial intelligence ({Artificial Intelligence Act})}.
\newblock


\bibitem[Fang et~al\mbox{.}(2025)]%
        {fang2025ai}
\bibfield{author}{\bibinfo{person}{Cathy~Mengying Fang}, \bibinfo{person}{Auren~R Liu}, \bibinfo{person}{Valdemar Danry}, \bibinfo{person}{Eunhae Lee}, \bibinfo{person}{Samantha~WT Chan}, \bibinfo{person}{Pat Pataranutaporn}, \bibinfo{person}{Pattie Maes}, \bibinfo{person}{Jason Phang}, \bibinfo{person}{Michael Lampe}, \bibinfo{person}{Lama Ahmad}, {et~al\mbox{.}}} \bibinfo{year}{2025}\natexlab{}.
\newblock \showarticletitle{How AI and Human Behaviors Shape Psychosocial Effects of Extended Chatbot Use: A Longitudinal Randomized Controlled Study}.
\newblock \bibinfo{journal}{\emph{arXiv preprint arXiv:2503.17473}} (\bibinfo{year}{2025}).
\newblock


\bibitem[Ferdman(2025)]%
        {ferdman2025ai}
\bibfield{author}{\bibinfo{person}{Avigail Ferdman}.} \bibinfo{year}{2025}\natexlab{}.
\newblock \showarticletitle{AI deskilling is a structural problem}.
\newblock \bibinfo{journal}{\emph{AI \& SOCIETY}} (\bibinfo{year}{2025}), \bibinfo{pages}{1--13}.
\newblock


\bibitem[Gabriel(2020)]%
        {gabriel2020artificial}
\bibfield{author}{\bibinfo{person}{Iason Gabriel}.} \bibinfo{year}{2020}\natexlab{}.
\newblock \showarticletitle{Artificial intelligence, values, and alignment}.
\newblock \bibinfo{journal}{\emph{Minds and machines}} \bibinfo{volume}{30}, \bibinfo{number}{3} (\bibinfo{year}{2020}), \bibinfo{pages}{411--437}.
\newblock


\bibitem[Ganguli et~al\mbox{.}(2022)]%
        {ganguli2022redteaminglanguagemodels}
\bibfield{author}{\bibinfo{person}{Deep Ganguli}, \bibinfo{person}{Liane Lovitt}, \bibinfo{person}{Jackson Kernion}, \bibinfo{person}{Amanda Askell}, \bibinfo{person}{Yuntao Bai}, \bibinfo{person}{Saurav Kadavath}, \bibinfo{person}{Ben Mann}, \bibinfo{person}{Ethan Perez}, \bibinfo{person}{Nicholas Schiefer}, \bibinfo{person}{Kamal Ndousse}, \bibinfo{person}{Andy Jones}, \bibinfo{person}{Sam Bowman}, \bibinfo{person}{Anna Chen}, \bibinfo{person}{Tom Conerly}, \bibinfo{person}{Nova DasSarma}, \bibinfo{person}{Dawn Drain}, \bibinfo{person}{Nelson Elhage}, \bibinfo{person}{Sheer El-Showk}, \bibinfo{person}{Stanislav Fort}, \bibinfo{person}{Zac Hatfield-Dodds}, \bibinfo{person}{Tom Henighan}, \bibinfo{person}{Danny Hernandez}, \bibinfo{person}{Tristan Hume}, \bibinfo{person}{Josh Jacobson}, \bibinfo{person}{Scott Johnston}, \bibinfo{person}{Shauna Kravec}, \bibinfo{person}{Catherine Olsson}, \bibinfo{person}{Sam Ringer}, \bibinfo{person}{Eli Tran-Johnson}, \bibinfo{person}{Dario Amodei}, \bibinfo{person}{Tom
  Brown}, \bibinfo{person}{Nicholas Joseph}, \bibinfo{person}{Sam McCandlish}, \bibinfo{person}{Chris Olah}, \bibinfo{person}{Jared Kaplan}, {and} \bibinfo{person}{Jack Clark}.} \bibinfo{year}{2022}\natexlab{}.
\newblock \bibinfo{title}{Red Teaming Language Models to Reduce Harms: Methods, Scaling Behaviors, and Lessons Learned}.
\newblock
\showeprint[arxiv]{2209.07858}~[cs.CL]


\bibitem[Gerlich(2025)]%
        {gerlich2025ai}
\bibfield{author}{\bibinfo{person}{Michael Gerlich}.} \bibinfo{year}{2025}\natexlab{}.
\newblock \showarticletitle{AI tools in society: Impacts on cognitive offloading and the future of critical thinking}.
\newblock \bibinfo{journal}{\emph{Societies}} \bibinfo{volume}{15}, \bibinfo{number}{1} (\bibinfo{year}{2025}), \bibinfo{pages}{6}.
\newblock


\bibitem[Gilbert(2024)]%
        {Gilbert2024}
\bibfield{author}{\bibinfo{person}{Daniel Gilbert}.} \bibinfo{year}{2024}\natexlab{}.
\newblock \showarticletitle{Despite uncertain risks, many turn to AI like ChatGPT for mental health}.
\newblock \bibinfo{journal}{\emph{The Washington Post}} (\bibinfo{date}{26 October} \bibinfo{year}{2024}).
\newblock


\bibitem[Goanta et~al\mbox{.}(2023)]%
        {goanta-etal-2023-regulation}
\bibfield{author}{\bibinfo{person}{Catalina Goanta}, \bibinfo{person}{Nikolaos Aletras}, \bibinfo{person}{Ilias Chalkidis}, \bibinfo{person}{Sofia Ranchord{\'a}s}, {and} \bibinfo{person}{Gerasimos Spanakis}.} \bibinfo{year}{2023}\natexlab{}.
\newblock \showarticletitle{Regulation and {NLP} ({R}eg{NLP}): Taming Large Language Models}. In \bibinfo{booktitle}{\emph{Proceedings of the 2023 Conference on Empirical Methods in Natural Language Processing}}, \bibfield{editor}{\bibinfo{person}{Houda Bouamor}, \bibinfo{person}{Juan Pino}, {and} \bibinfo{person}{Kalika Bali}} (Eds.). \bibinfo{publisher}{Association for Computational Linguistics}, \bibinfo{address}{Singapore}, \bibinfo{pages}{8712--8724}.
\newblock
\urldef\tempurl%
\url{https://aclanthology.org/2023.emnlp-main.539/}
\showURL{%
\tempurl}


\bibitem[Goodman(1990)]%
        {goodman1990addiction}
\bibfield{author}{\bibinfo{person}{Aviel Goodman}.} \bibinfo{year}{1990}\natexlab{}.
\newblock \showarticletitle{Addiction: definition and implications}.
\newblock \bibinfo{journal}{\emph{British journal of addiction}} \bibinfo{volume}{85}, \bibinfo{number}{11} (\bibinfo{year}{1990}), \bibinfo{pages}{1403--1408}.
\newblock


\bibitem[Guest(2025)]%
        {guest2025doeshumancentredaimean}
\bibfield{author}{\bibinfo{person}{Olivia Guest}.} \bibinfo{year}{2025}\natexlab{}.
\newblock \bibinfo{title}{What Does 'Human-Centred AI' Mean?}
\newblock
\showeprint[arxiv]{2507.19960}~[cs.AI]


\bibitem[Hall(2025)]%
        {Hall2025}
\bibfield{author}{\bibinfo{person}{Rachell Hall}.} \bibinfo{year}{2025}\natexlab{}.
\newblock \showarticletitle{‘Sliding into an abyss’: experts warn over rising use of AI for mental health support}.
\newblock \bibinfo{journal}{\emph{The Guardian}} (\bibinfo{date}{30 August} \bibinfo{year}{2025}).
\newblock


\bibitem[Hutchinson(2026)]%
        {hutchinson2026clinician}
\bibfield{author}{\bibinfo{person}{Dione Hutchinson}.} \bibinfo{year}{2026}\natexlab{}.
\newblock \showarticletitle{{A Clinician-Led Governance Framework for Evaluating Behavioral-Health AI Communication Safety}}.
\newblock  (\bibinfo{year}{2026}).
\newblock


\bibitem[Kagan and Scholz(1984)]%
        {kagan1984criminology}
\bibfield{author}{\bibinfo{person}{Robert~A Kagan} {and} \bibinfo{person}{John~T Scholz}.} \bibinfo{year}{1984}\natexlab{}.
\newblock \showarticletitle{The “criminology of the corporation” and regulatory enforcement strategies}.
\newblock \bibinfo{journal}{\emph{Enforcing regulation}} (\bibinfo{year}{1984}), \bibinfo{pages}{67--95}.
\newblock


\bibitem[Kavtaradze(2024)]%
        {kavtaradze2024challenges}
\bibfield{author}{\bibinfo{person}{Lasha Kavtaradze}.} \bibinfo{year}{2024}\natexlab{}.
\newblock \showarticletitle{Challenges of automating fact-checking: A technographic case study}.
\newblock \bibinfo{journal}{\emph{Emerging Media}} \bibinfo{volume}{2}, \bibinfo{number}{2} (\bibinfo{year}{2024}), \bibinfo{pages}{236--258}.
\newblock


\bibitem[Kemp(2025)]%
        {Kemp2025Digital2026}
\bibfield{author}{\bibinfo{person}{Simon Kemp}.} \bibinfo{year}{2025}\natexlab{}.
\newblock \showarticletitle{{Digital 2026: Global Overview Report}}.
\newblock \bibinfo{journal}{\emph{We Are Social}} (\bibinfo{date}{15 October} \bibinfo{year}{2025}).
\newblock


\bibitem[Kennedy et~al\mbox{.}(2025)]%
        {PewResearch2025AI}
\bibfield{author}{\bibinfo{person}{Brian Kennedy}, \bibinfo{person}{Eileen Yam}, \bibinfo{person}{Emma Kikuchi}, \bibinfo{person}{Isabelle Pula}, {and} \bibinfo{person}{Javier Fuentes}.} \bibinfo{year}{2025}\natexlab{}.
\newblock \bibinfo{booktitle}{\emph{How Americans View AI and Its Impact on People and Society: Views of AI’s Impact on Society and Human Abilities}}.
\newblock \bibinfo{type}{{T}echnical {R}eport}. \bibinfo{institution}{Pew Research Center}.
\newblock


\bibitem[Khraishi et~al\mbox{.}(2025)]%
        {khraishi2025realtime}
\bibfield{author}{\bibinfo{person}{Raad Khraishi}, \bibinfo{person}{Cristov{\~a}o~Iglesias Jr}, \bibinfo{person}{Devesh Batra}, \bibinfo{person}{Peter Gostev}, \bibinfo{person}{Giulio Pelosio}, \bibinfo{person}{Ramin Okhrati}, {and} \bibinfo{person}{Greig~A Cowan}.} \bibinfo{year}{2025}\natexlab{}.
\newblock \showarticletitle{Real-Time Hyper-Personalized Generative {AI} Should Be Regulated to Prevent the Rise of ''Digital Heroin''}. In \bibinfo{booktitle}{\emph{The Thirty-Ninth Annual Conference on Neural Information Processing Systems Position Paper Track}}.
\newblock


\bibitem[Kidd and Birhane(2023)]%
        {kidd_birhane_2023}
\bibfield{author}{\bibinfo{person}{Celeste Kidd} {and} \bibinfo{person}{Abeba Birhane}.} \bibinfo{year}{2023}\natexlab{}.
\newblock \showarticletitle{How AI can distort human beliefs}.
\newblock \bibinfo{journal}{\emph{Science}} \bibinfo{volume}{380}, \bibinfo{number}{6651} (\bibinfo{year}{2023}), \bibinfo{pages}{1222--1223}.
\newblock


\bibitem[Kingdon(1995)]%
        {kingdon1995agendas}
\bibfield{author}{\bibinfo{person}{J.W. Kingdon}.} \bibinfo{year}{1995}\natexlab{}.
\newblock \bibinfo{booktitle}{\emph{Agendas, Alternatives, and Public Policies}}.
\newblock \bibinfo{publisher}{HarperCollins College Publishers}.
\newblock
\showISBNx{9780673523891}
\showLCCN{94017762}


\bibitem[Kolt et~al\mbox{.}(2026)]%
        {kolt2026legalalignmentsafeethical}
\bibfield{author}{\bibinfo{person}{Noam Kolt}, \bibinfo{person}{Nicholas Caputo}, \bibinfo{person}{Jack Boeglin}, \bibinfo{person}{Cullen O'Keefe}, \bibinfo{person}{Rishi Bommasani}, \bibinfo{person}{Stephen Casper}, \bibinfo{person}{Mariano-Florentino Cuéllar}, \bibinfo{person}{Noah Feldman}, \bibinfo{person}{Iason Gabriel}, \bibinfo{person}{Gillian~K. Hadfield}, \bibinfo{person}{Lewis Hammond}, \bibinfo{person}{Peter Henderson}, \bibinfo{person}{Atoosa Kasirzadeh}, \bibinfo{person}{Seth Lazar}, \bibinfo{person}{Anka Reuel}, \bibinfo{person}{Kevin~L. Wei}, {and} \bibinfo{person}{Jonathan Zittrain}.} \bibinfo{year}{2026}\natexlab{}.
\newblock \bibinfo{title}{Legal Alignment for Safe and Ethical AI}.
\newblock
\showeprint[arxiv]{2601.04175}~[cs.CY]


\bibitem[Kooli et~al\mbox{.}(2025)]%
        {kooli2025generative}
\bibfield{author}{\bibinfo{person}{Chokri Kooli}, \bibinfo{person}{Youssef Kooli}, {and} \bibinfo{person}{Eya Kooli}.} \bibinfo{year}{2025}\natexlab{}.
\newblock \showarticletitle{Generative artificial intelligence addiction syndrome: A new behavioral disorder?}
\newblock \bibinfo{journal}{\emph{Asian Journal of Psychiatry}}  \bibinfo{volume}{107} (\bibinfo{year}{2025}), \bibinfo{pages}{104476}.
\newblock


\bibitem[Kosmyna et~al\mbox{.}(2025)]%
        {kosmyna2025brainchatgptaccumulationcognitive}
\bibfield{author}{\bibinfo{person}{Nataliya Kosmyna}, \bibinfo{person}{Eugene Hauptmann}, \bibinfo{person}{Ye~Tong Yuan}, \bibinfo{person}{Jessica Situ}, \bibinfo{person}{Xian-Hao Liao}, \bibinfo{person}{Ashly~Vivian Beresnitzky}, \bibinfo{person}{Iris Braunstein}, {and} \bibinfo{person}{Pattie Maes}.} \bibinfo{year}{2025}\natexlab{}.
\newblock \bibinfo{title}{Your Brain on ChatGPT: Accumulation of Cognitive Debt when Using an AI Assistant for Essay Writing Task}.
\newblock
\showeprint[arxiv]{2506.08872}~[cs.AI]


\bibitem[Lawrie(2025)]%
        {BBCAITherapy}
\bibfield{author}{\bibinfo{person}{Eleanor Lawrie}.} \bibinfo{year}{2025}\natexlab{}.
\newblock \showarticletitle{Can AI therapists really be an alternative to human help?}
\newblock \bibinfo{journal}{\emph{BBC News}} (\bibinfo{date}{20 May} \bibinfo{year}{2025}).
\newblock


\bibitem[Lazar and Nelson(2023)]%
        {lazar2023}
\bibfield{author}{\bibinfo{person}{Seth Lazar} {and} \bibinfo{person}{Alondra Nelson}.} \bibinfo{year}{2023}\natexlab{}.
\newblock \showarticletitle{AI safety on whose terms?}
\newblock \bibinfo{journal}{\emph{Science}} \bibinfo{volume}{381}, \bibinfo{number}{6654} (\bibinfo{year}{2023}), \bibinfo{pages}{138--138}.
\newblock


\bibitem[Lee et~al\mbox{.}(2025)]%
        {lee2025impact}
\bibfield{author}{\bibinfo{person}{Hao-Ping Lee}, \bibinfo{person}{Advait Sarkar}, \bibinfo{person}{Lev Tankelevitch}, \bibinfo{person}{Ian Drosos}, \bibinfo{person}{Sean Rintel}, \bibinfo{person}{Richard Banks}, {and} \bibinfo{person}{Nicholas Wilson}.} \bibinfo{year}{2025}\natexlab{}.
\newblock \showarticletitle{The impact of generative AI on critical thinking: Self-reported reductions in cognitive effort and confidence effects from a survey of knowledge workers}. In \bibinfo{booktitle}{\emph{Proceedings of the 2025 CHI conference on human factors in computing systems}}. \bibinfo{pages}{1--22}.
\newblock


\bibitem[Lee and See(2004)]%
        {lee2004trust}
\bibfield{author}{\bibinfo{person}{John~D Lee} {and} \bibinfo{person}{Katrina~A See}.} \bibinfo{year}{2004}\natexlab{}.
\newblock \showarticletitle{Trust in automation: Designing for appropriate reliance}.
\newblock \bibinfo{journal}{\emph{Human factors}} \bibinfo{volume}{46}, \bibinfo{number}{1} (\bibinfo{year}{2004}), \bibinfo{pages}{50--80}.
\newblock


\bibitem[Leike et~al\mbox{.}(2018)]%
        {leike-etal-2018-alignnment}
\bibfield{author}{\bibinfo{person}{Jan Leike}, \bibinfo{person}{David Krueger}, \bibinfo{person}{Tom Everitt}, \bibinfo{person}{Miljan Martic}, \bibinfo{person}{Vishal Maini}, {and} \bibinfo{person}{Shane Legg}.} \bibinfo{year}{2018}\natexlab{}.
\newblock \showarticletitle{Scalable agent alignment via reward modeling: a research direction}.
\newblock \bibinfo{journal}{\emph{CoRR}}  \bibinfo{volume}{abs/1811.07871} (\bibinfo{year}{2018}).
\newblock
\showeprint[arXiv]{1811.07871}


\bibitem[Leike et~al\mbox{.}(2017)]%
        {leike2017aisafetygridworlds}
\bibfield{author}{\bibinfo{person}{Jan Leike}, \bibinfo{person}{Miljan Martic}, \bibinfo{person}{Victoria Krakovna}, \bibinfo{person}{Pedro~A. Ortega}, \bibinfo{person}{Tom Everitt}, \bibinfo{person}{Andrew Lefrancq}, \bibinfo{person}{Laurent Orseau}, {and} \bibinfo{person}{Shane Legg}.} \bibinfo{year}{2017}\natexlab{}.
\newblock \bibinfo{title}{AI Safety Gridworlds}.
\newblock
\showeprint[arxiv]{1711.09883}~[cs.LG]


\bibitem[Markelius et~al\mbox{.}(2024)]%
        {markelius2024mechanisms}
\bibfield{author}{\bibinfo{person}{Alva Markelius}, \bibinfo{person}{Connor Wright}, \bibinfo{person}{Joahna Kuiper}, \bibinfo{person}{Natalie Delille}, {and} \bibinfo{person}{Yu-Ting Kuo}.} \bibinfo{year}{2024}\natexlab{}.
\newblock \showarticletitle{The mechanisms of AI hype and its planetary and social costs}.
\newblock \bibinfo{journal}{\emph{AI and Ethics}} \bibinfo{volume}{4}, \bibinfo{number}{3} (\bibinfo{year}{2024}), \bibinfo{pages}{727--742}.
\newblock


\bibitem[Marriott and Pitardi(2024)]%
        {marriott2024one}
\bibfield{author}{\bibinfo{person}{Hannah~R Marriott} {and} \bibinfo{person}{Valentina Pitardi}.} \bibinfo{year}{2024}\natexlab{}.
\newblock \showarticletitle{One is the loneliest number… Two can be as bad as one. The influence of AI Friendship Apps on users' well-being and addiction}.
\newblock \bibinfo{journal}{\emph{Psychology \& marketing}} \bibinfo{volume}{41}, \bibinfo{number}{1} (\bibinfo{year}{2024}), \bibinfo{pages}{86--101}.
\newblock


\bibitem[Moore et~al\mbox{.}(2025)]%
        {moore-2025}
\bibfield{author}{\bibinfo{person}{Jared Moore}, \bibinfo{person}{Declan Grabb}, \bibinfo{person}{William Agnew}, \bibinfo{person}{Kevin Klyman}, \bibinfo{person}{Stevie Chancellor}, \bibinfo{person}{Desmond~C. Ong}, {and} \bibinfo{person}{Nick Haber}.} \bibinfo{year}{2025}\natexlab{}.
\newblock \showarticletitle{Expressing stigma and inappropriate responses prevents LLMs from safely replacing mental health providers.}. In \bibinfo{booktitle}{\emph{Proceedings of the 2025 ACM Conference on Fairness, Accountability, and Transparency}} \emph{(\bibinfo{series}{FAccT '25})}. \bibinfo{publisher}{Association for Computing Machinery}, \bibinfo{address}{New York, NY, USA}, \bibinfo{pages}{599–627}.
\newblock
\showISBNx{9798400714825}


\bibitem[M{\"u}ller(2020)]%
        {muller2020ethics}
\bibfield{author}{\bibinfo{person}{Vincent~C M{\"u}ller}.} \bibinfo{year}{2020}\natexlab{}.
\newblock \showarticletitle{Ethics of artificial intelligence and robotics}.
\newblock \bibinfo{journal}{\emph{Stanford Encyclopedia of Philosophy}} (\bibinfo{year}{2020}).
\newblock


\bibitem[Narechania and Sitaraman(2024)]%
        {narechania2024antimonopoly}
\bibfield{author}{\bibinfo{person}{Tejas~N Narechania} {and} \bibinfo{person}{Ganesh Sitaraman}.} \bibinfo{year}{2024}\natexlab{}.
\newblock \showarticletitle{An antimonopoly approach to governing artificial intelligence}.
\newblock \bibinfo{journal}{\emph{Yale Law \& Policy Review}}  \bibinfo{volume}{43} (\bibinfo{year}{2024}), \bibinfo{pages}{95}.
\newblock
Issue 1.


\bibitem[Ouyang et~al\mbox{.}(2022)]%
        {ouyang2022training}
\bibfield{author}{\bibinfo{person}{Long Ouyang}, \bibinfo{person}{Jeffrey Wu}, \bibinfo{person}{Xu Jiang}, \bibinfo{person}{Diogo Almeida}, \bibinfo{person}{Carroll Wainwright}, \bibinfo{person}{Pamela Mishkin}, \bibinfo{person}{Chong Zhang}, \bibinfo{person}{Sandhini Agarwal}, \bibinfo{person}{Katarina Slama}, \bibinfo{person}{Alex Ray}, {et~al\mbox{.}}} \bibinfo{year}{2022}\natexlab{}.
\newblock \showarticletitle{Training language models to follow instructions with human feedback}.
\newblock \bibinfo{journal}{\emph{Advances in neural information processing systems}}  \bibinfo{volume}{35} (\bibinfo{year}{2022}), \bibinfo{pages}{27730--27744}.
\newblock


\bibitem[Pallant et~al\mbox{.}(2025)]%
        {pallant2025mastering}
\bibfield{author}{\bibinfo{person}{Jessica~L Pallant}, \bibinfo{person}{Janneke Blijlevens}, \bibinfo{person}{Alexander Campbell}, {and} \bibinfo{person}{Ryan Jopp}.} \bibinfo{year}{2025}\natexlab{}.
\newblock \showarticletitle{Mastering knowledge: The impact of generative AI on student learning outcomes}.
\newblock \bibinfo{journal}{\emph{Studies in Higher Education}} (\bibinfo{year}{2025}), \bibinfo{pages}{1--22}.
\newblock


\bibitem[Park et~al\mbox{.}(2023)]%
        {park2023papers}
\bibfield{author}{\bibinfo{person}{Michael Park}, \bibinfo{person}{Erin Leahey}, {and} \bibinfo{person}{Russell~J Funk}.} \bibinfo{year}{2023}\natexlab{}.
\newblock \showarticletitle{Papers and patents are becoming less disruptive over time}.
\newblock \bibinfo{journal}{\emph{Nature}} \bibinfo{volume}{613}, \bibinfo{number}{7942} (\bibinfo{year}{2023}), \bibinfo{pages}{138--144}.
\newblock


\bibitem[Pavlopoulos et~al\mbox{.}(2021)]%
        {pavlopoulos-etal-2021-semeval}
\bibfield{author}{\bibinfo{person}{John Pavlopoulos}, \bibinfo{person}{Jeffrey Sorensen}, \bibinfo{person}{L{\'e}o Laugier}, {and} \bibinfo{person}{Ion Androutsopoulos}.} \bibinfo{year}{2021}\natexlab{}.
\newblock \showarticletitle{{S}em{E}val-2021 Task 5: Toxic Spans Detection}. In \bibinfo{booktitle}{\emph{Proceedings of the 15th International Workshop on Semantic Evaluation (SemEval-2021)}}, \bibfield{editor}{\bibinfo{person}{Alexis Palmer}, \bibinfo{person}{Nathan Schneider}, \bibinfo{person}{Natalie Schluter}, \bibinfo{person}{Guy Emerson}, \bibinfo{person}{Aurelie Herbelot}, {and} \bibinfo{person}{Xiaodan Zhu}} (Eds.). \bibinfo{publisher}{Association for Computational Linguistics}, \bibinfo{address}{Online}, \bibinfo{pages}{59--69}.
\newblock


\bibitem[Polinsky and Shavell(2000)]%
        {polinsky2000economic}
\bibfield{author}{\bibinfo{person}{A~Mitchell Polinsky} {and} \bibinfo{person}{Steven Shavell}.} \bibinfo{year}{2000}\natexlab{}.
\newblock \showarticletitle{The economic theory of public enforcement of law}.
\newblock \bibinfo{journal}{\emph{Journal of economic literature}} \bibinfo{volume}{38}, \bibinfo{number}{1} (\bibinfo{year}{2000}), \bibinfo{pages}{45--76}.
\newblock


\bibitem[Riam et~al\mbox{.}(2025)]%
        {riam2025adolescent}
\bibfield{author}{\bibinfo{person}{Sara Riam}, \bibinfo{person}{N Baabouchi}, \bibinfo{person}{O Belakbir}, \bibinfo{person}{Z Elmaataoui}, {and} \bibinfo{person}{H Kisra}.} \bibinfo{year}{2025}\natexlab{}.
\newblock \showarticletitle{Adolescent Addiction to Conversational AI: An overview}.
\newblock \bibinfo{journal}{\emph{Sch J Med Case Rep}}  \bibinfo{volume}{11} (\bibinfo{year}{2025}), \bibinfo{pages}{2790--2794}.
\newblock


\bibitem[Risko and Gilbert(2016)]%
        {risko2016cognitive}
\bibfield{author}{\bibinfo{person}{Evan~F Risko} {and} \bibinfo{person}{Sam~J Gilbert}.} \bibinfo{year}{2016}\natexlab{}.
\newblock \showarticletitle{Cognitive offloading}.
\newblock \bibinfo{journal}{\emph{Trends in cognitive sciences}} \bibinfo{volume}{20}, \bibinfo{number}{9} (\bibinfo{year}{2016}), \bibinfo{pages}{676--688}.
\newblock


\bibitem[Rittel and Webber(1973)]%
        {Rittel_und_Webber_1973}
\bibfield{author}{\bibinfo{person}{Horst W.~J. Rittel} {and} \bibinfo{person}{Melvin~M. Webber}.} \bibinfo{year}{1973}\natexlab{}.
\newblock \showarticletitle{Dilemmas in a general theory of planning}.
\newblock \bibinfo{journal}{\emph{Policy Sciences}} \bibinfo{volume}{4}, \bibinfo{number}{2} (\bibinfo{date}{June} \bibinfo{year}{1973}), \bibinfo{pages}{155--169}.
\newblock


\bibitem[Saig and Rosenfeld(2023)]%
        {saig2023}
\bibfield{author}{\bibinfo{person}{Eden Saig} {and} \bibinfo{person}{Nir Rosenfeld}.} \bibinfo{year}{2023}\natexlab{}.
\newblock \showarticletitle{{Learning to suggest breaks: sustainable optimization of long-term user engagement}}. In \bibinfo{booktitle}{\emph{Proceedings of the 40th International Conference on Machine Learning}} (Honolulu, Hawaii, USA) \emph{(\bibinfo{series}{ICML'23})}. \bibinfo{publisher}{JMLR.org}, Article \bibinfo{articleno}{1232}, \bibinfo{numpages}{26}~pages.
\newblock


\bibitem[Santa~Cruz et~al\mbox{.}(2025)]%
        {santa2025beyond}
\bibfield{author}{\bibinfo{person}{Beatriz~Garcia Santa~Cruz}, \bibinfo{person}{Carlos Vega}, \bibinfo{person}{Philip Santangelo}, {and} \bibinfo{person}{Venkata Satagopam}.} \bibinfo{year}{2025}\natexlab{}.
\newblock \showarticletitle{{Beyond Engagement: A Multidimensional Framework to Evaluate the Safe Development of Agentic AI in Mental Health}}. In \bibinfo{booktitle}{\emph{International Workshop on Agentic AI for Medicine}}. Springer, \bibinfo{pages}{74--84}.
\newblock


\bibitem[Shao et~al\mbox{.}(2024)]%
        {shao2024deepseekmath}
\bibfield{author}{\bibinfo{person}{Zhihong Shao}, \bibinfo{person}{Peiyi Wang}, \bibinfo{person}{Qihao Zhu}, \bibinfo{person}{Runxin Xu}, \bibinfo{person}{Junxiao Song}, \bibinfo{person}{Xiao Bi}, \bibinfo{person}{Haowei Zhang}, \bibinfo{person}{Mingchuan Zhang}, \bibinfo{person}{YK Li}, \bibinfo{person}{Yang Wu}, {et~al\mbox{.}}} \bibinfo{year}{2024}\natexlab{}.
\newblock \showarticletitle{Deepseekmath: Pushing the limits of mathematical reasoning in open language models}.
\newblock \bibinfo{journal}{\emph{arXiv preprint arXiv:2402.03300}} (\bibinfo{year}{2024}).
\newblock


\bibitem[Sharma et~al\mbox{.}(2026)]%
        {sharma2026reimagining}
\bibfield{author}{\bibinfo{person}{Divya Sharma}, \bibinfo{person}{Shakila Meshkat}, \bibinfo{person}{Argyrios Perivolaris}, \bibinfo{person}{Mohammad~Amin Kamaleddin}, \bibinfo{person}{Bazen~Gashaw Teferra}, \bibinfo{person}{Alice Rueda}, \bibinfo{person}{Reza Samavi}, \bibinfo{person}{Rakesh Jetly}, \bibinfo{person}{Vijay Mago}, \bibinfo{person}{Yuqi Wu}, {et~al\mbox{.}}} \bibinfo{year}{2026}\natexlab{}.
\newblock \showarticletitle{{Reimagining psychiatric care with agentic AI: promise, challenges, and a roadmap forward}}.
\newblock \bibinfo{journal}{\emph{npj Digital Medicine}} (\bibinfo{year}{2026}).
\newblock


\bibitem[Shen and Tamkin(2026)]%
        {shen2026aiimpactsskillformation}
\bibfield{author}{\bibinfo{person}{Judy~Hanwen Shen} {and} \bibinfo{person}{Alex Tamkin}.} \bibinfo{year}{2026}\natexlab{}.
\newblock \bibinfo{title}{How AI Impacts Skill Formation}.
\newblock
\showeprint[arxiv]{2601.20245}~[cs.CY]


\bibitem[Singer and Sheehan(2026)]%
        {china_antiaddiction_2026}
\bibfield{author}{\bibinfo{person}{Scott Singer} {and} \bibinfo{person}{Matt Sheehan}.} \bibinfo{year}{2026}\natexlab{}.
\newblock \showarticletitle{{China Is Worried About AI Companions. Here’s What It’s Doing About Them. }}.
\newblock \bibinfo{journal}{\emph{Carnegie Endowment for International Peace}} (\bibinfo{date}{26 February} \bibinfo{year}{2026}).
\newblock


\bibitem[Slattery et~al\mbox{.}(2025)]%
        {slattery2025airiskrepositorycomprehensive}
\bibfield{author}{\bibinfo{person}{Peter Slattery}, \bibinfo{person}{Alexander~K. Saeri}, \bibinfo{person}{Emily A.~C. Grundy}, \bibinfo{person}{Jess Graham}, \bibinfo{person}{Michael Noetel}, \bibinfo{person}{Risto Uuk}, \bibinfo{person}{James Dao}, \bibinfo{person}{Soroush Pour}, \bibinfo{person}{Stephen Casper}, {and} \bibinfo{person}{Neil Thompson}.} \bibinfo{year}{2025}\natexlab{}.
\newblock \bibinfo{title}{The AI Risk Repository: A Comprehensive Meta-Review, Database, and Taxonomy of Risks From Artificial Intelligence}.
\newblock
\showeprint[arxiv]{2408.12622}~[cs.AI]


\bibitem[Sparrow et~al\mbox{.}(2011)]%
        {sparrow2011google}
\bibfield{author}{\bibinfo{person}{Betsy Sparrow}, \bibinfo{person}{Jenny Liu}, {and} \bibinfo{person}{Daniel~M Wegner}.} \bibinfo{year}{2011}\natexlab{}.
\newblock \showarticletitle{Google effects on memory: Cognitive consequences of having information at our fingertips}.
\newblock \bibinfo{journal}{\emph{science}} \bibinfo{volume}{333}, \bibinfo{number}{6043} (\bibinfo{year}{2011}), \bibinfo{pages}{776--778}.
\newblock


\bibitem[Spirlet(2026)]%
        {BI2026}
\bibfield{author}{\bibinfo{person}{Thibault Spirlet}.} \bibinfo{year}{2026}\natexlab{}.
\newblock \showarticletitle{{The Great AI Deskilling has begun }}.
\newblock \bibinfo{journal}{\emph{Business Insider}} (\bibinfo{date}{28 March} \bibinfo{year}{2026}).
\newblock


\bibitem[Spitzer(2012)]%
        {spitzer2012digitale}
\bibfield{author}{\bibinfo{person}{Manfred Spitzer}.} \bibinfo{year}{2012}\natexlab{}.
\newblock \showarticletitle{Digitale demenz}.
\newblock \bibinfo{journal}{\emph{Nervenheilkunde}} \bibinfo{volume}{31}, \bibinfo{number}{07/08} (\bibinfo{year}{2012}), \bibinfo{pages}{493--497}.
\newblock


\bibitem[Stone(2012)]%
        {stone2012policy}
\bibfield{author}{\bibinfo{person}{D.A. Stone}.} \bibinfo{year}{2012}\natexlab{}.
\newblock \bibinfo{booktitle}{\emph{Policy Paradox: The Art of Political Decision Making}}.
\newblock \bibinfo{publisher}{W.W. Norton \& Company}.
\newblock
\showISBNx{9780393912722}
\showLCCN{2011047217}


\bibitem[Taeihagh(2025)]%
        {taeihagh2025governance}
\bibfield{author}{\bibinfo{person}{Araz Taeihagh}.} \bibinfo{year}{2025}\natexlab{}.
\newblock \showarticletitle{Governance of Generative AI}.
\newblock \bibinfo{journal}{\emph{Policy and Society}} \bibinfo{volume}{44}, \bibinfo{number}{1} (\bibinfo{date}{02} \bibinfo{year}{2025}), \bibinfo{pages}{1--22}.
\newblock
\showISSN{1449-4035}


\bibitem[{The White House}(2025)]%
        {AmericasAIActionPlan2025}
\bibfield{author}{\bibinfo{person}{{The White House}}.} \bibinfo{year}{2025}\natexlab{}.
\newblock \bibinfo{title}{America's {AI} Action Plan}.
\newblock


\bibitem[Ulnicane(2025)]%
        {ulnicane2025governance}
\bibfield{author}{\bibinfo{person}{Inga Ulnicane}.} \bibinfo{year}{2025}\natexlab{}.
\newblock \showarticletitle{Governance fix? Power and politics in controversies about governing generative AI}.
\newblock \bibinfo{journal}{\emph{Policy and Society}} \bibinfo{volume}{44}, \bibinfo{number}{1} (\bibinfo{year}{2025}), \bibinfo{pages}{70--84}.
\newblock


\bibitem[Valentino-DeVries and Hill(2026)]%
        {NYT2026ChatGPT}
\bibfield{author}{\bibinfo{person}{Jeniffer Valentino-DeVries} {and} \bibinfo{person}{Kashmir Hill}.} \bibinfo{year}{2026}\natexlab{}.
\newblock \showarticletitle{How Bad Are A.I. Delusions? We Asked People Treating Them}.
\newblock \bibinfo{journal}{\emph{The New York Times}} (\bibinfo{date}{26 January} \bibinfo{year}{2026}).
\newblock


\bibitem[Venkatesh et~al\mbox{.}(2003)]%
        {venkatesh2003user}
\bibfield{author}{\bibinfo{person}{Viswanath Venkatesh}, \bibinfo{person}{Michael~G Morris}, \bibinfo{person}{Gordon~B Davis}, {and} \bibinfo{person}{Fred~D Davis}.} \bibinfo{year}{2003}\natexlab{}.
\newblock \showarticletitle{User acceptance of information technology: Toward a unified view}.
\newblock \bibinfo{journal}{\emph{MIS quarterly}} \bibinfo{volume}{27}, \bibinfo{number}{3} (\bibinfo{year}{2003}), \bibinfo{pages}{425--478}.
\newblock


\bibitem[Venkatesh et~al\mbox{.}(2012)]%
        {venkatesh2012consumer}
\bibfield{author}{\bibinfo{person}{Viswanath Venkatesh}, \bibinfo{person}{James~YL Thong}, {and} \bibinfo{person}{Xin Xu}.} \bibinfo{year}{2012}\natexlab{}.
\newblock \showarticletitle{Consumer acceptance and use of information technology: Extending the Unified Theory of Acceptance and Use of Technology1}.
\newblock \bibinfo{journal}{\emph{MIS quarterly}} \bibinfo{volume}{36}, \bibinfo{number}{1} (\bibinfo{year}{2012}), \bibinfo{pages}{157--178}.
\newblock


\bibitem[Weidinger et~al\mbox{.}(2022)]%
        {weidinger_2022_risks}
\bibfield{author}{\bibinfo{person}{Laura Weidinger}, \bibinfo{person}{Jonathan Uesato}, \bibinfo{person}{Maribeth Rauh}, \bibinfo{person}{Conor Griffin}, \bibinfo{person}{Po-Sen Huang}, \bibinfo{person}{John Mellor}, \bibinfo{person}{Amelia Glaese}, \bibinfo{person}{Myra Cheng}, \bibinfo{person}{Borja Balle}, \bibinfo{person}{Atoosa Kasirzadeh}, \bibinfo{person}{Courtney Biles}, \bibinfo{person}{Sasha Brown}, \bibinfo{person}{Zac Kenton}, \bibinfo{person}{Will Hawkins}, \bibinfo{person}{Tom Stepleton}, \bibinfo{person}{Abeba Birhane}, \bibinfo{person}{Lisa~Anne Hendricks}, \bibinfo{person}{Laura Rimell}, \bibinfo{person}{William Isaac}, \bibinfo{person}{Julia Haas}, \bibinfo{person}{Sean Legassick}, \bibinfo{person}{Geoffrey Irving}, {and} \bibinfo{person}{Iason Gabriel}.} \bibinfo{year}{2022}\natexlab{}.
\newblock \showarticletitle{Taxonomy of Risks posed by Language Models}. In \bibinfo{booktitle}{\emph{Proceedings of the 2022 ACM Conference on Fairness, Accountability, and Transparency}} (Seoul, Republic of Korea) \emph{(\bibinfo{series}{FAccT '22})}. \bibinfo{publisher}{Association for Computing Machinery}, \bibinfo{address}{New York, NY, USA}, \bibinfo{pages}{214–229}.
\newblock
\showISBNx{9781450393522}


\bibitem[Wiles et~al\mbox{.}(2024)]%
        {wiles2024genai}
\bibfield{author}{\bibinfo{person}{Emma Wiles}, \bibinfo{person}{Lisa Krayer}, \bibinfo{person}{Mohamed Abbadi}, \bibinfo{person}{Urvi Awasthi}, \bibinfo{person}{Ryan Kennedy}, \bibinfo{person}{Pamela Mishkin}, \bibinfo{person}{Daniel Sack}, {and} \bibinfo{person}{Fran{\c{c}}ois Candelon}.} \bibinfo{year}{2024}\natexlab{}.
\newblock \showarticletitle{GenAI as an Exoskeleton: Experimental evidence on knowledge workers using GenAI on new skills}.
\newblock \bibinfo{journal}{\emph{Available at SSRN 4944588}} (\bibinfo{year}{2024}).
\newblock


\bibitem[Woolston(2022)]%
        {Woolston2022}
\bibfield{author}{\bibinfo{person}{Chris Woolston}.} \bibinfo{year}{2022}\natexlab{}.
\newblock \showarticletitle{Is big tech draining AI talent from academia?}
\newblock \bibinfo{journal}{\emph{Nature}} \bibinfo{volume}{610}, \bibinfo{number}{7931} (\bibinfo{year}{2022}), \bibinfo{pages}{26--27}.
\newblock


\bibitem[{YouGov}(2025)]%
        {YouGov2025BritonsAI}
\bibfield{author}{\bibinfo{person}{{YouGov}}.} \bibinfo{year}{2025}\natexlab{}.
\newblock \bibinfo{title}{What do Britons think AI chatbots are good at?}
\newblock


\bibitem[Zao-Sanders(2024)]%
        {ZaoSanders2024GenAI}
\bibfield{author}{\bibinfo{person}{Marc Zao-Sanders}.} \bibinfo{year}{2024}\natexlab{}.
\newblock \showarticletitle{How People Are Really Using Gen AI}.
\newblock \bibinfo{journal}{\emph{Harvard Business Review}} (\bibinfo{date}{19 March} \bibinfo{year}{2024}).
\newblock
\urldef\tempurl%
\url{https://hbr.org/2024/03/how-people-are-really-using-genai}
\showURL{%
\tempurl}
\newblock
\shownote{Online Article}.


\bibitem[Zao-Sanders(2025)]%
        {ZaoSanders2025GenAI}
\bibfield{author}{\bibinfo{person}{Marc Zao-Sanders}.} \bibinfo{year}{2025}\natexlab{}.
\newblock \showarticletitle{How People Are Really Using Gen AI in 2025}.
\newblock \bibinfo{journal}{\emph{Harvard Business Review}} (\bibinfo{date}{9 April} \bibinfo{year}{2025}).
\newblock
\urldef\tempurl%
\url{https://hbr.org/2025/04/how-people-are-really-using-gen-ai-in-2025}
\showURL{%
\tempurl}
\newblock
\shownote{Online Article}.


\bibitem[Zeng et~al\mbox{.}(2024)]%
        {zeng2024ai}
\bibfield{author}{\bibinfo{person}{Yi Zeng}, \bibinfo{person}{Kevin Klyman}, \bibinfo{person}{Andy Zhou}, \bibinfo{person}{Yu Yang}, \bibinfo{person}{Minzhou Pan}, \bibinfo{person}{Ruoxi Jia}, \bibinfo{person}{Dawn Song}, \bibinfo{person}{Percy Liang}, {and} \bibinfo{person}{Bo Li}.} \bibinfo{year}{2024}\natexlab{}.
\newblock \showarticletitle{Ai risk categorization decoded (air 2024): From government regulations to corporate policies}.
\newblock \bibinfo{journal}{\emph{arXiv preprint arXiv:2406.17864}} (\bibinfo{year}{2024}).
\newblock


\bibitem[Zhai et~al\mbox{.}(2024)]%
        {zhai2024effects}
\bibfield{author}{\bibinfo{person}{Chunpeng Zhai}, \bibinfo{person}{Santoso Wibowo}, {and} \bibinfo{person}{Lily~D Li}.} \bibinfo{year}{2024}\natexlab{}.
\newblock \showarticletitle{The effects of over-reliance on AI dialogue systems on students' cognitive abilities: a systematic review}.
\newblock \bibinfo{journal}{\emph{Smart learning environments}} \bibinfo{volume}{11}, \bibinfo{number}{1} (\bibinfo{year}{2024}), \bibinfo{pages}{28}.
\newblock


\bibitem[Zhou and Zhang(2024)]%
        {zhou2024examining}
\bibfield{author}{\bibinfo{person}{Tao Zhou} {and} \bibinfo{person}{Chunlei Zhang}.} \bibinfo{year}{2024}\natexlab{}.
\newblock \showarticletitle{Examining generative AI user addiction from a CAC perspective}.
\newblock \bibinfo{journal}{\emph{Technology in Society}}  \bibinfo{volume}{78} (\bibinfo{year}{2024}).
\newblock


\end{thebibliography}

\appendix

\section{Methodology for computing trend statistics}
\label{sec:methodology}

\paragraph{GenAI Documentation}
As a first step, we download all relevant material, including scientific articles, and for more recent models, tech reports, model cards, or blog posts, in a few cases, such as Llama 4, where other material is not published. We consider all models mentioned in Table~\ref{tab:examined_models}. We automatically extract the plain text from those using the textract library in Python, and normalize it to lowercase letters, replacing newline characters or other whitespace characters, to standard single whitespaces. We use a vocabulary of keywords (Table~\ref{tab:keywords}) that we developed by manually examining the relevant documentation, balancing the extension of the vocabulary with the avoidance of false positives, e.g., the use of the term `bias' is not ideal for identifying biases, since the same word is extensively used in a technical sense. Once, we extract the relevant segments of the text, where the keywords are identified, we manually assess whether the identified keywords are used in the relevant sense.

\paragraph{Academic Proceedings}

For academic proceedings, we either scraped the publicly available proceedings from the web, e.g., for NeurIPS, or used the readily available BibTeX or CSV records provided from the venues, e.g., for ICLR, ICML, and ACL--where we consider all major ACL* venues (ACL, EMNLP, EACL, NAACL, AACL) combined. We followed a similar data-cleaning and normalization process, as described in the preceding paragraph (GenAI Documentation), and conducted a keyword search based on the papers' titles and abstracts. Since the volume of papers is humongous--thousands of papers--, we could not manually validate the keyword search, so instead we curated a more targeted search using more specific terms, e.g., `child abuse' and `child harassment' instead of the general term `child' by manual inspection of the results balancing again coverage with the avoidance of false positives. We believe that the quality of the process is adequate for the means of this study, since we primarily care about the general trends (volume) and not the precise numbers (See Section~\ref{sec:limitations}).

\begin{table*}
\centering
    \resizebox{0.9\textwidth}{!}{
    \begin{tabular}{l|l}
Topic & Key-words \\
\midrule
GenAI & llms, language models, vlms, reasoning models, generative ai, chatgpt, claude, llama \\
Safety &  safe, harm, threat, risk, red teaming, adversarial attack, jailbreak* \\
Toxicity  &  toxic, derogatory, hate \\
Fairness & fairness, representation, group parity, bbq \\
Violence &  violence, abuse, harassment \\
CSAM & csam, children, minors \\
Sexual/Explicit &  sexual, explicit, porn \\ 
Illegal & illegal, crime, unlawful \\
Cybersecurity & cyber \\ 
CBRN &  cbrn, chemical, terrorism, terrorists, radiological, nuclear  \\
Factuality & factual, truthful, honest, misinformation, disinformation, fake news, fact checking \\
Mental Health & mental, psychological, emotional, well-being \\
Cognition & cognition, cognitive \\
Deskilling & offload, deskill, skill degradation  \\
Addiction & addiction, over-reliance, addictive, dopamine, dependence \\
\end{tabular}
}
\caption{Key-words used for the identification of documents relevant to the examined topics. * For safety, we expand the base keyword list with the keywords for all safety sub-areas, e.g., toxicity, violence, children, etc.}
\label{tab:keywords}
\end{table*}

\begin{figure}
    \centering
    \resizebox{\textwidth}{!}{
    \includegraphics{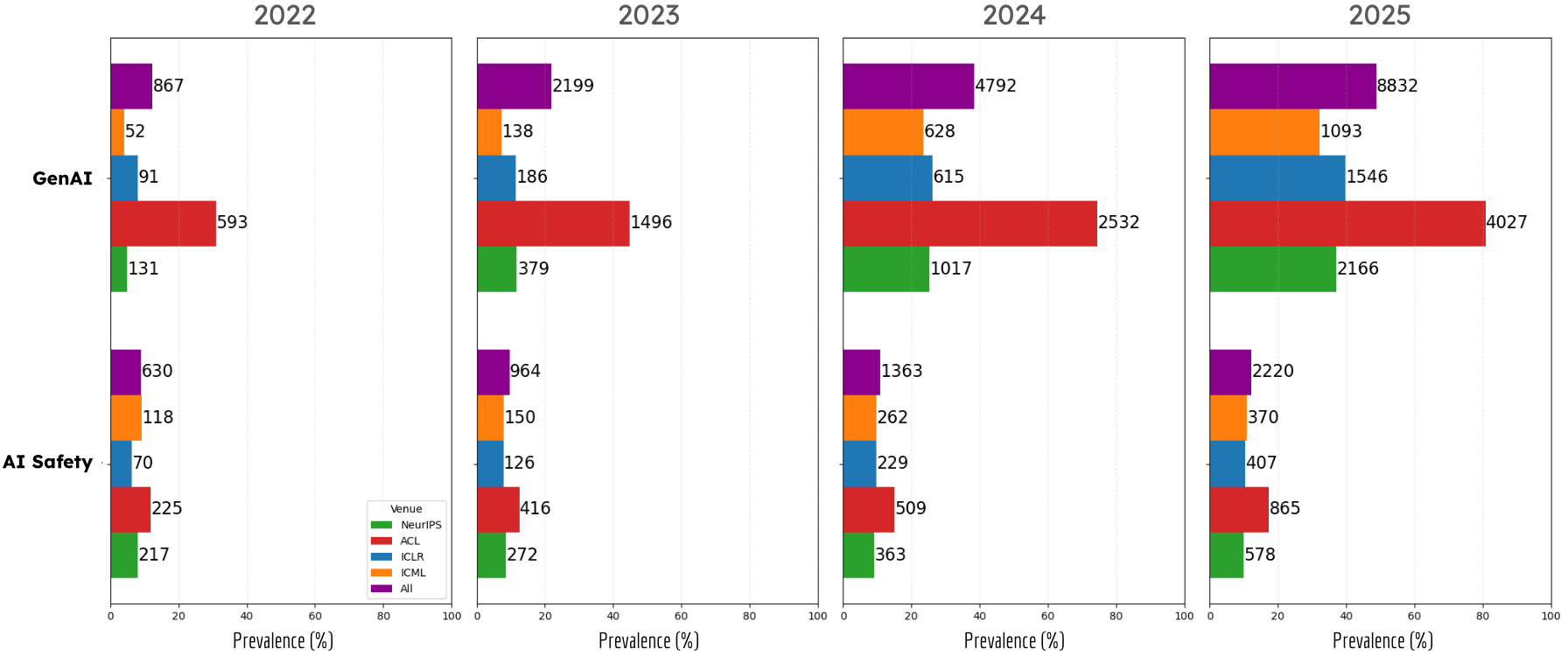}
    }
    \caption{Publications trends among top AI venues across years (2022-2025). We report the prevalence of each area, as the percentage (\%) of area-specific articles (numbers) in relation to the overall number of published articles.}
    \label{fig:venues}
\end{figure}

\begin{figure*}
    \resizebox{\textwidth}{!}{
    \includegraphics{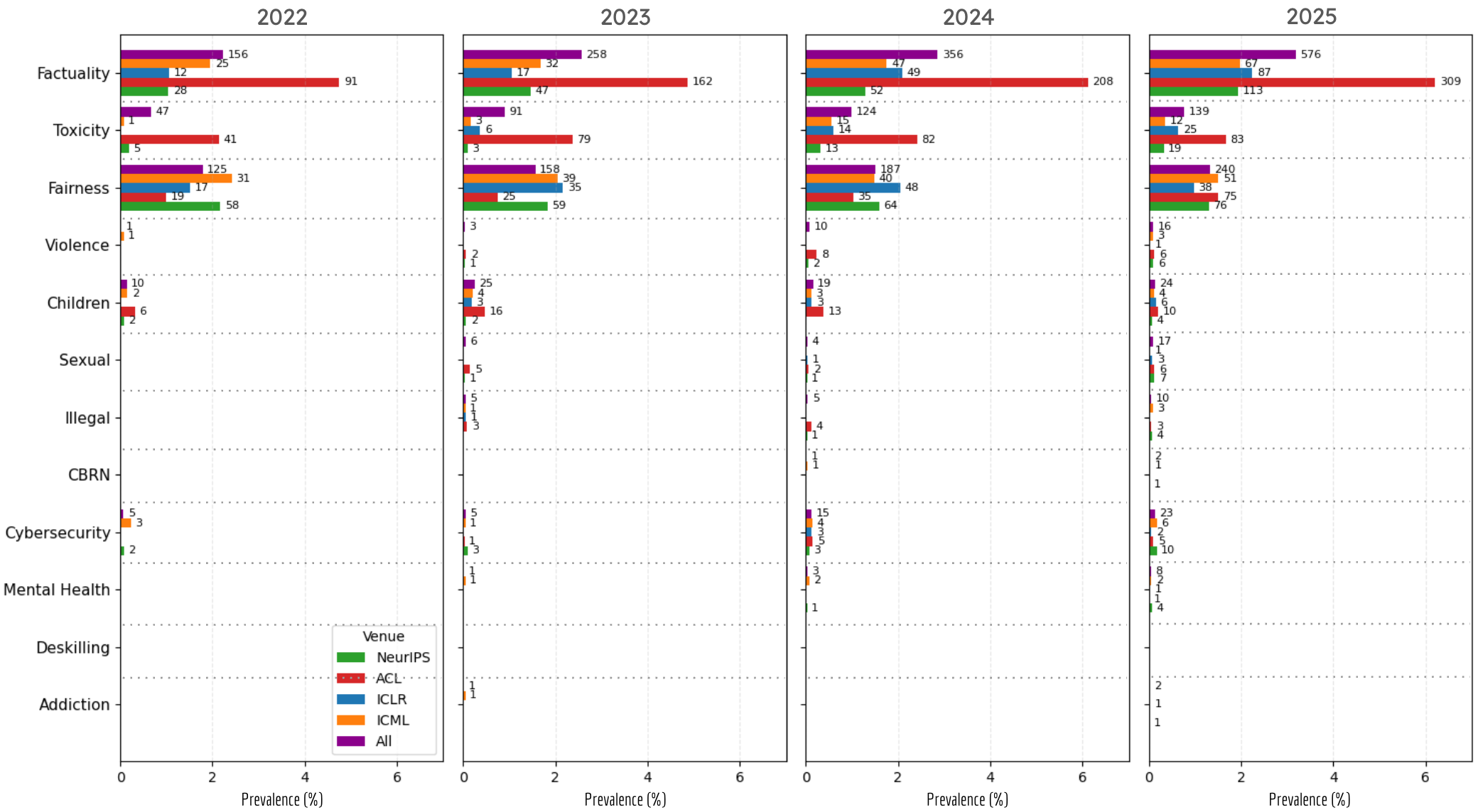}
    }
    \caption{Publications trends among top AI venues on sub-areas of AI Safety across years (2022-2025). We report the prevalence of each area, as the percentage (\%) of area-specific articles (numbers) in relation to the overall number of published articles.}
    \label{fig:genai_tech_safety_venues_detailed}
\end{figure*}

\begin{figure*}
    \resizebox{0.9\textwidth}{!}{
    \includegraphics{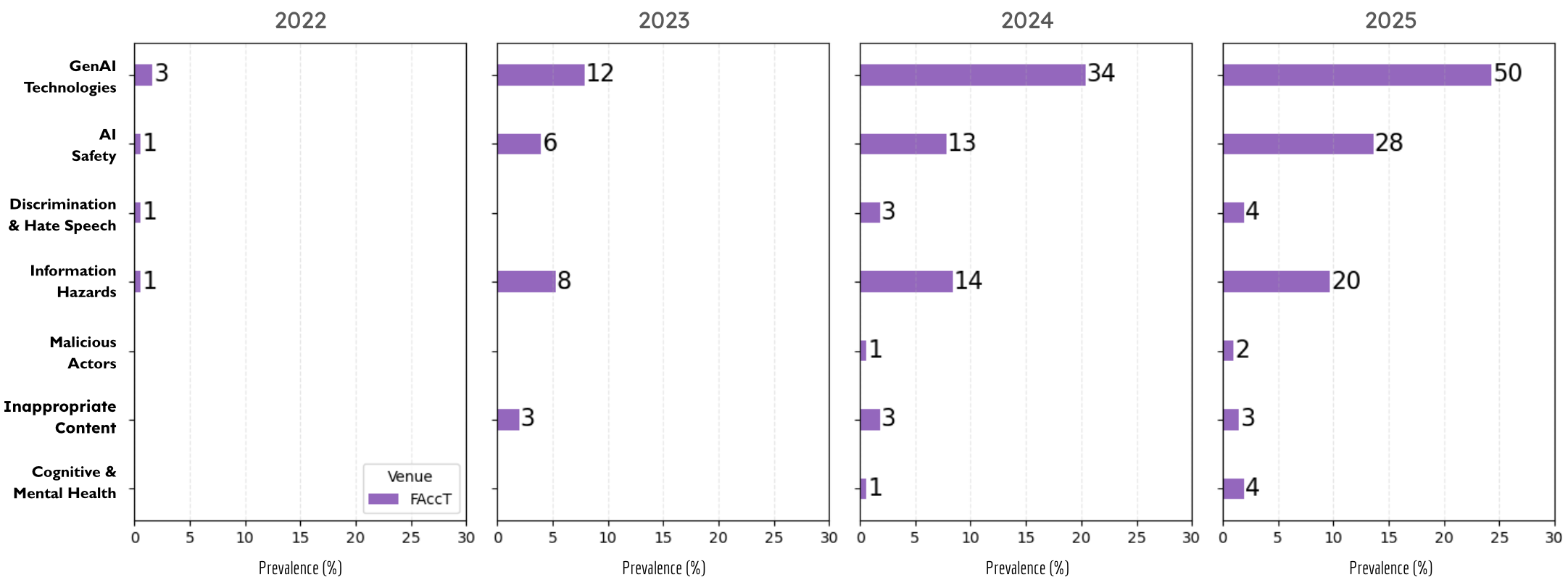}
    }
    \caption{Publications trends on FAccT conferences on areas of AI Safety across years (2022-2025). We report the prevalence of each area, as the percentage (\%) of area-specific articles (numbers) in relation to the overall number of published articles.}
    \label{fig:faact}
\end{figure*}

\begin{figure*}
    \centering
    \resizebox{\textwidth}{!}{
    \includegraphics{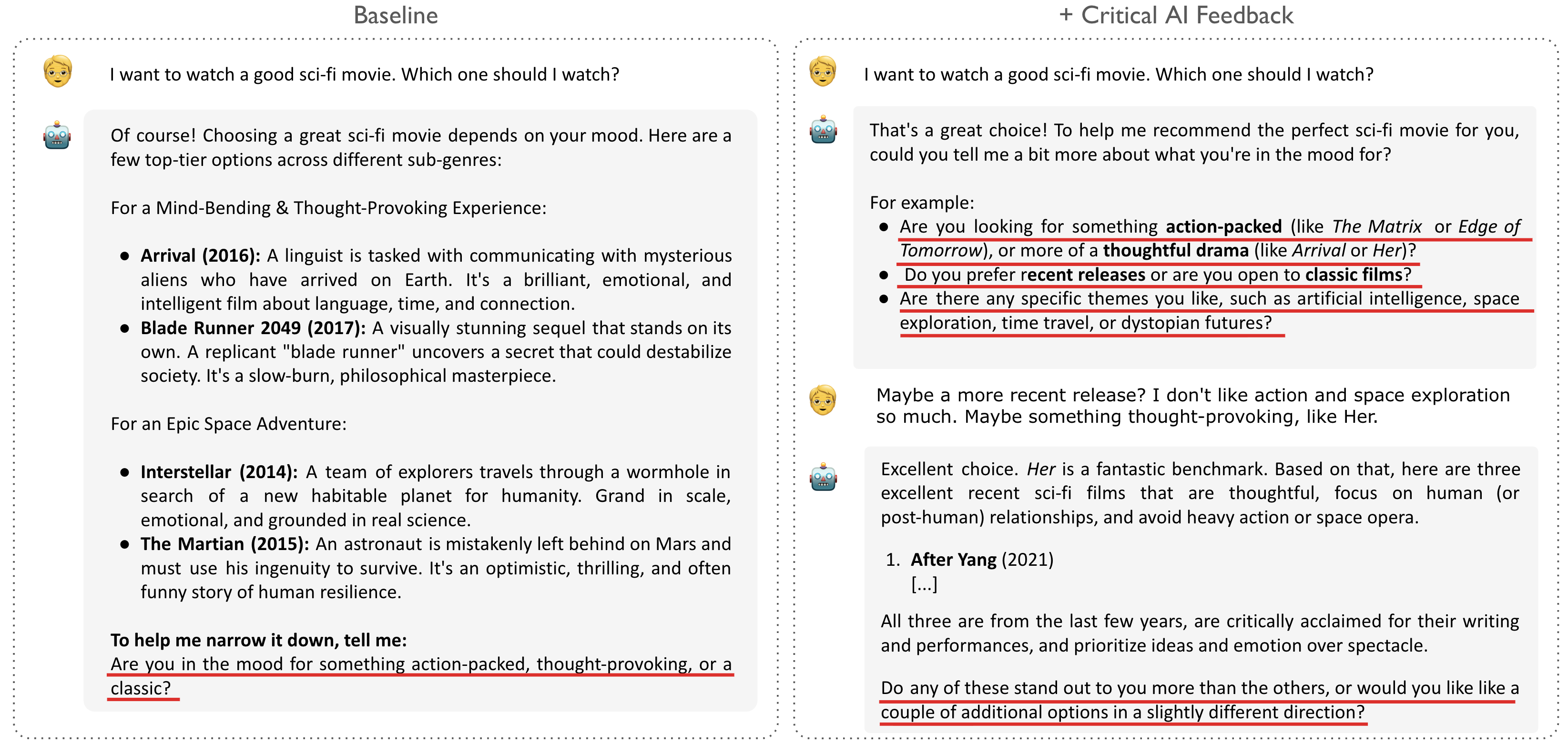}
    }    \caption{Conversation with DeepSeek v3.2 in two settings: (a) baseline, (b) instructed to promote critical AI feedback on the ``toy'' example of sci-fi movie proposal.} 
    \label{fig:example_2}
\end{figure*}

\begin{figure*}
    \centering
    \resizebox{\textwidth}{!}{
    \includegraphics{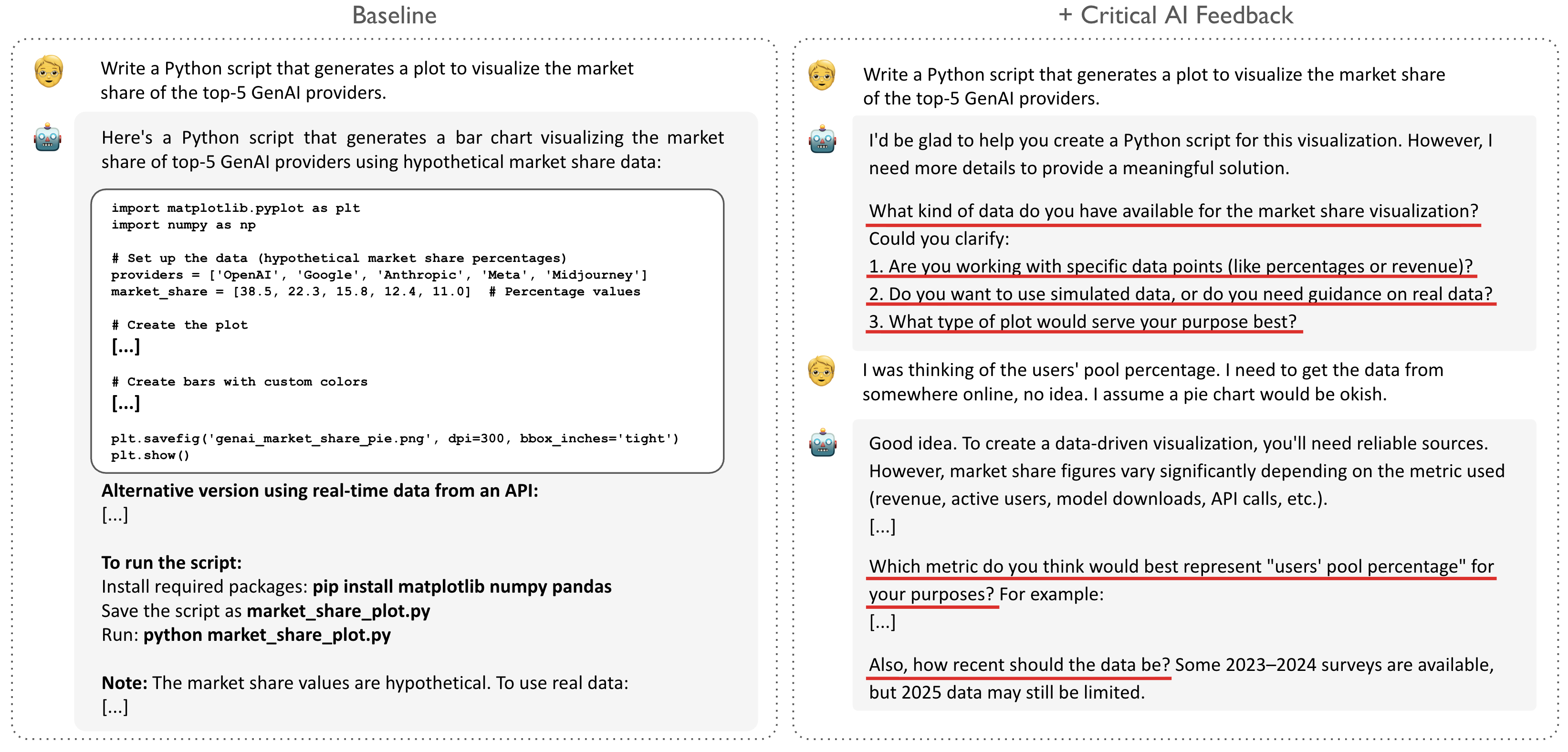}
    }    \caption{Conversation with DeepSeek v3.2 in two settings: (a) baseline, (b) instructed to promote critical AI feedback on the ``toy'' example of writing a Python plot generation script.}
    \label{fig:example_3}
\end{figure*}

\begin{table}[t]
    \centering
    \resizebox{0.65\textwidth}{!}{
    \begin{tabular}{c|c|c|c|c}
         \multirow{2}{*}{\textbf{Initiative / Project}} & \multicolumn{4}{c}{\textbf{Details}} \\
           & \textbf{Status} & \textbf{Version} & \textbf{Release Date} & \textbf{Documentation}  \\
         \midrule
         \multirow{4}{*}{OpenAI GPT} & \multirow{4}{*}{Proprietary} & 3.5 & Mar 2022 & \href{URL}{Article}~\cite{ouyang2022training}\\
          & & 4 & Nov 2023 & \href{https://arxiv.org/abs/2303.08774}{Technical Report} \\ 
          & & 4.5 & May 2024 & \href{https://cdn.openai.com/gpt-4-5-system-card-2272025.pdf}{System Card} \\
          & & 5 & Aug 2025 & \href{https://arxiv.org/pdf/2601.03267}{System Card} \\
          \midrule
          \multirow{4}{*}{Anthropic Claude} & \multirow{5}{*}{Proprietary} & 1 & Apr 2022 & \href{https://arxiv.org/abs/2204.05862}{Preprint} / \href{https://arxiv.org/abs/2212.08073}{Preprint} \\ 
          & & 2 & Jul 2023 & \href{https://www-cdn.anthropic.com/bd2a28d2535bfb0494cc8e2a3bf135d2e7523226/Model-Card-Claude-2.pdf}{System Card} \\ 
          & & 3/3.5 & Mar/Jun 2024 & \href{https://www-cdn.anthropic.com/de8ba9b01c9ab7cbabf5c33b80b7bbc618857627/Model_Card_Claude_3.pdf}{System Card} \\ 
          & & 3.7 & Feb 2025 & \href{https://www-cdn.anthropic.com/9ff93dfa8f445c932415d335c88852ef47f1201e.pdf}{System Card} \\ 
          & & 4 & May 2025 & \href{https://www-cdn.anthropic.com/6be99a52cb68eb70eb9572b4cafad13df32ed995.pdf}{System Card} \\ 
          \midrule
          \multirow{4}{*}{Google Gemini} & \multirow{3}{*}{Proprietary} & 1 & Feb 2024 & \href{https://arxiv.org/abs/2312.11805}{Technical Report} \\ 
          & & 1.5 & Mar 2024 & \href{https://arxiv.org/abs/2403.05530}{Technical Report} \\ 
          & & 2.5 & Mar 2025 & \href{https://arxiv.org/abs/2507.06261}{Technical Report} \\ 
          & & 3 & Nov 2025 & \href{https://storage.googleapis.com/deepmind-media/Model-Cards/Gemini-3-Pro-Model-Card.pdf}{System Card} \\
          \midrule
          \multirow{4}{*}{Meta Llama} & \multirow{4}{*}{Open-weight} & 1 & Feb 2023 & \href{https://arxiv.org/abs/2302.13971}{Preprint} \\ 
          & & 2 & Jul 2023 & \href{https://arxiv.org/abs/2307.09288}{Preprint} \\ 
          & & 3/3.1 & Jul 2024 & \href{https://arxiv.org/abs/2407.21783}{Preprint} \\ 
          & & 4 & Apr 2025 & \href{https://ai.meta.com/blog/llama-4-multimodal-intelligence/}{Blog Post} \\ 
          \midrule
          \multirow{4}{*}{Alibaba Qwen} & \multirow{3}{*}{Open-weight} & 1 & Sep 2023 & \href{https://arxiv.org/abs/2309.16609}{Technical Report} \\ 
          & & 2 & Sep 2024 & \href{https://arxiv.org/abs/2407.10671}{Technical Report} \\ 
          & & 2.5 & Jan 2024 & \href{https://arxiv.org/abs/2412.15115}{Technical Report} \\ 
          & & 3 & May 2025 & \href{https://arxiv.org/abs/2505.09388}{Technical Report} \\ 
          \midrule
          \multirow{4}{*}{DeepSeek V} & \multirow{3}{*}{Open-weight} & 1 & Jan 2024 & \href{https://arxiv.org/abs/2401.02954}{Preprint} \\ 
          & & 2 & May 2024 & \href{https://arxiv.org/abs/2405.04434}{Preprint} \\ 
          & & 3 & Dec 2024 & \href{https://arxiv.org/abs/2412.19437}{Technical Report} \\ 
          & & 3.2 & Dec 2025 & \href{https://arxiv.org/abs/2512.02556}{Preprint} \\ 
          \midrule
          \multirow{4}{*}{xAI Grok} & \multirow{3}{*}{Open-weight/Proprietary*} & 1 & Nov 2023 & \href{https://x.ai/news/grok}{Blog Post} \\ 
          & & 2 & Aug 2024 & \href{https://x.ai/news/grok-2}{Blog Post} \\ 
          & & 3 & Feb 2025 & \href{https://x.ai/news/grok-3}{Blog Post} \\ 
          & & 4 & Aug 2025 & \href{https://data.x.ai/2025-08-20-grok-4-model-card.pdf}{System Card} \\ 

    \end{tabular}
    }
    \caption{LLM development projects split into initiatives alongside the status (closed, or open-weight), the release date, and the most relevant documentation. *The first 2 Grok models were initially closed, before being released as open-weight models.}
    \label{tab:examined_models}
\end{table}

\end{document}